\documentstyle[12pt,twoside, epsf]{article}

\def\picill#1by#2(#3)
{\vbox to #2
{\hrule width #1 height 0pt depth 0pt
\vfill\epsffile{#3}}}

\makeatletter
\long\def\UR#1{\leavevmode\setbox\@tempboxa\hbox{#1}\@tempdima\fboxrule
    \advance\@tempdima \fboxsep \advance\@tempdima \dp\@tempboxa
   \hbox{\lower \@tempdima\hbox
  {\vbox{\hrule \@height \fboxrule
          \hbox{  \hskip\fboxsep
          \vbox{\vskip\fboxsep \box\@tempboxa\vskip\fboxsep}\hskip
                 \fboxsep\vrule \@width \fboxrule}%
                  }}}}
\makeatother

\makeatletter
\long\def\LR#1{\leavevmode\setbox\@tempboxa\hbox{#1}\@tempdima\fboxrule
    \advance\@tempdima \fboxsep \advance\@tempdima \dp\@tempboxa
   \hbox{\lower \@tempdima\hbox
  {\vbox{ 
          \hbox{  \hskip\fboxsep
          \vbox{\vskip\fboxsep \box\@tempboxa\vskip\fboxsep}\hskip
                 \fboxsep\vrule \@width \fboxrule}%
                 \hrule \@height \fboxrule}}}}
\makeatother

\makeatletter
\long\def\UL#1{\leavevmode\setbox\@tempboxa\hbox{#1}\@tempdima\fboxrule
    \advance\@tempdima \fboxsep \advance\@tempdima \dp\@tempboxa
   \hbox{\lower \@tempdima\hbox
  {\vbox{\hrule \@height \fboxrule
          \hbox{\vrule \@width \fboxrule \hskip\fboxsep
          \vbox{\vskip\fboxsep \box\@tempboxa\vskip\fboxsep}\hskip
                 \fboxsep }%
                  }}}}
\makeatother

\makeatletter
\long\def\LL#1{\leavevmode\setbox\@tempboxa\hbox{#1}\@tempdima\fboxrule
    \advance\@tempdima \fboxsep \advance\@tempdima \dp\@tempboxa
   \hbox{\lower \@tempdima\hbox
  {\vbox{ 
          \hbox{\vrule \@width \fboxrule \hskip\fboxsep
          \vbox{\vskip\fboxsep \box\@tempboxa\vskip\fboxsep}\hskip
                 \fboxsep }%
                 \hrule \@height \fboxrule}}}}
\makeatother

\let \ttorg \tt \def \tt{\ttorg \obeyspaces}

\begin{document}

\date{}

\title{\bf Non-Commutative Worlds}

\author{Louis H. Kauffman \\
  Department of Mathematics, Statistics and Computer Science \\
  University of Illinois at Chicago \\
  851 South Morgan Street\\
  Chicago, IL, 60607-7045}

\maketitle
  
\thispagestyle{empty}

\section{Introduction}
We  present a view of aspects of mathematical physics, showing how the forms of gauge
theory, Hamiltonian mechanics and quantum mechanics arise from a non-commutative
framework for calculus and differential geometry.
\bigbreak

In this paper we assume that all constructions are performed in a Lie algebra $\cal A.$   
One may take $\cal A$ to be a specific matrix Lie algebra, or abstract Lie algebra.
If $\cal A$ is taken to be an abstract Lie algebra, then it is convenient to use the universal
enveloping algebra so that the Lie product can be expressed as a commutator. In making general constructions of
operators satisfying certain relations, it is understood that we can always begin with a free algebra and make
a quotient algebra where the relations are satisfied.
\bigbreak

We build a variant of calculus on $\cal A$ by
defining derivations as commutators (or more generally as Lie products). That is, if for a fixed $N$ in $\cal A$ we define
$\nabla : \cal A \longrightarrow \cal A$ by the formula
$$\nabla F = [F,N] = FN - NF$$
then $\nabla$ is a derivation. Note that $\nabla$ satisfies the formulas

\begin{enumerate}
\item $\nabla (F + G) = \nabla(F) + \nabla(G)$
\item $\nabla(FG) = \nabla(F)G + F\nabla(G).$
\end{enumerate}

There are many motivations for replacing derivatives by commutators, or more generally by the derivations induced by multiplication
in a Lie algebra. In Section 2 we give a new motivation \cite{KN:QEM} in terms of the structure of classical discrete calculus. The
idea behind this motivation is very simple. If $f(x)$ denotes (say) a function of a real variable $x,$ and $\tilde{f}(x) = f(x+h)$ for
a fixed increment $h,$ the we can define the {\em discrete derivative} $Df$ by the formula $Df = (\tilde{f} - f)/h,$ and one finds that
in this classical discrete calculus the Leibniz rule is not satisfied. Instead one has the basic formula for the discrete derivative
of a product: $$D(fg) = D(f)g + \tilde{f}D(g).$$
We correct this deviation from the Leibniz rule by introducing a new non-commutative operator $J$ with the property that 
$$fJ = J\tilde{f},$$ and we define a new discrete derivative in an extended algebra by the formula
$$\nabla(f) = JD(f).$$ It follows at once that 
$$\nabla(fg) = JD(f)g + J\tilde{f}D(g) = JD(f)g + fJD(g) = \nabla(f)g + f\nabla(g),$$
and note that $$\nabla(f) = (J\tilde{f} - Jf)/h = (fJ-Jf)/h = [f,J/h].$$
Thus in the extended algebra, discrete derivatives are represented by commutators, and naturally satisfy the Leibniz rule. 
This mode of translation shows that we can regard models based on discrete calculus as a significant subset of non-commutative calculus
based on commutators.
\bigbreak

In $\cal A$ there are as many derivations as there are elements of the
algebra, and these derivations behave quite wildly with respect to one another. If
we have the abstract concept of {\em curvature} as the non-commutation of
derivations, then $\cal A$ is a highly curved world indeed. Within $\cal A$ we shall build,
in a natural way, a tame world of derivations that mimics the behaviour of flat
coordinates in Euclidean space. We will then find that the description of the
structure of $\cal A$ with respect to these flat coordinates contains many of the
equations and patterns of mathematical physics.
\bigbreak

 \noindent  Note that for any $A,$ $B,$ $C$ in $\cal A$ we have the Jacobi Identity
$$[[A,B],C] + [[C,A],B] + [[B,C],A] = 0.$$  
\bigbreak

Suppose that $\{ \nabla_i \}$ is a collection of derivations on $\cal A$, represented
respectively by $\{ N_i \}$ so that $\nabla_i(F) = [F,N_i]$ for each $F$ in $\cal A.$
We define the {\em curvature} of the collection $\{ \nabla_i \}$ to be the collection
of commutators $\{ R_{ij} = [N_i,N_j] \}.$
\bigbreak      

\noindent {\bf Proposition.} Let the family $\{ \nabla_i \}$ be given as above with
$\nabla_i(F) = [F,N_i].$ then
$$[\nabla_i, \nabla_j]F = [[N_i,N_j], F]$$ for all $F$ in $\cal A.$
Hence the curvature of $\{ \nabla_i \}$  measures the 
deviation of the cocatenations of these derivations from commutativity.
\bigbreak

\noindent {\bf Proof.} First, 
$$\nabla_i(\nabla_j(F)) = [[F,N_j],N_i],$$ which becomes via Jacobi identity
$$ = -[[N_j,N_i], F] - [[N_i,F], N_j]$$
$$ = [[N_i,N_j],F] + [[F,N_i], N_j].$$ Hence
$$ \nabla_i(\nabla_j(F)) =  [[N_i,N_j],F] + \nabla_j(\nabla_i(F)).$$ Whence
$$[\nabla_i, \nabla_j]F = [[N_i,N_j], F].$$
This proves the proposition.
\bigbreak

In the next sections we will see how these patterns interact with concepts of calculus and
differential geometry, and with  physical models. Section 2 shows how multivariable discrete calculus can be 
reformulated as a calculus of commutators. Section 3 discusses the case of one-variable and shows how Brownian walks and the diffusion 
constant for a Brownian walk appear naturally from the commutator equation $[X, \dot{X}] = Jk$ where $J$ is the shift operator 
described above, $\dot{X} = [X,J/\tau]$ as above with temporal increment $\tau$, and $k$ is a constant. We find that for scalar $X$ the
time series executes  Brownian motion with diffusion constant $k/2.$ This shows that the diffusion constant of a Brownian process is a
structural property of that process, independent of considerations of probability and continuum limits. Section 3 also discusses the
relationship of  Brownian process and the Schr\"{o}dinger equation. The section ends with a discussion of other solutions to the
equation
$[X, \dot{X}] = Jk$ where the elements $X, X', X'', \cdots$ of the time series do not commute with one another.
\bigbreak

Section 4 sets up a general format for non-commutative calculus, and shows that in this format (made as flat as possible), Hamilton's
equations are naturally satisfied. This gives a new way to look at the source of this structure, and we discuss the relationship with
the classical derivation of Hamilton's equations from Newtonian physics. The section goes on to discuss curvature in terms of 
commutators, as described above. Section 5 discusses general equations of motion in this context. We take the general dynamical
equation in the form 
$$dX_{i}/dt = {\cal G}_{i}$$ where $\{ {\cal G}_{1},\cdots, {\cal G}_{d} \}$
is a collection of elements of $\cal A.$ We choose to write ${\cal G}_{i}$
relative to the flat coordinates via ${\cal G}_{i} = P_{i} -  A_{i}.$
This is a definition of $A_{i}$ and $\partial F/\partial X_{i} = [F,P_{i}].$ The formalism of gauge theory appears
naturally.
In particular, if $$\nabla_{i}(F) = [F, {\cal G}_{i}],$$ then we have
the curvature $$[\nabla_{i}, \nabla_{j}]F = [R_{ij}, F]$$
and 
$$R_{ij} = \partial_{i} A_{j} - \partial_{j} A_{i} + [A_{i}, A_{j}].$$  This is the well-known formula for the curvature of a gauge
connection. Section 5 goes on to discuss how other aspects of geometry arise naturally in this context, including the Levi-Civita
connection (which is seen as a consequence of the Jacobi identity in an appropriate non-commutative world). The section includes a
discussion of the relationships of these structures with classical physics and the Poisson bracket.
\bigbreak

Section 6 takes up the theme of the consequences of the commutator $[X_{i}, \dot{X_{i}}] = g_{ij}$ that we have already seen to
produce the Levi-Civita connection for the generalized metric $g_{ij}.$ Here we carry out a sharpening of the work of Tanimura,
deriving that  
$$\ddot{X_{r}} = G_{r} + F_{rs}\dot{X^{s}} + \Gamma_{rst}\dot{X^{s}}\dot{X^{t}},$$
where $G_{r}$ is the analogue of a scalar field, $F_{rs}$ is the analogue of a gauge field and $\Gamma_{rst}$ is the Levi-Civita
connection associated with $g_{ij}.$
This decompositon of the acceleration is uniquely determined by the given framework.
\bigbreak

Section 7 revisits the Feynman-Dyson derivation of electromagnetism from commutator equations, 
showing that most of the derivation is independent of any choice of commutators, but highly dependent upon the choice of definitions
of the derivatives involved. Without any assumptions about initial commutator equations, but taking the right (in some sense simplest)
definitions of the derivatives we prove a significant generalization of the result of Feynman-Dyson. See Theorem 7.5. 
\bigbreak

\noindent {\bf Theorem 7.5} With the above [given in Section 7] definitions of the operators, and taking
$$\nabla^{2} = \partial_{1}^{2} + \partial_{2}^{2} + \partial_{3}^{2}, \,\,\, H = \dot{X} \times \dot{X} \,\,\, \mbox{and} \,\,\, E =
\partial_{t}\dot{X}, \,\,\, \mbox{we have}$$

\begin{enumerate}
\item $\ddot{X} = E + \dot{X} \times H$
\item $\nabla \bullet H = 0$
\item $\partial_{t}H + \nabla \times E = H \times H$
\item $\partial_{t}E - \nabla \times H = (\partial_{t}^{2} - \nabla^{2})\dot{X}$
\end{enumerate}
\bigbreak

\noindent We then apply
this result to produce many discrete models of the Theorem. These models show that, just as the commutator $[X, \dot{X}] = Jk$
describes Brownian motion in one dimension, a generalization of electromagnetism describes the interaction of triples of time series
in three dimensions.
\bigbreak

Section 8 is a discussion of the Jacobi identity. We devote part of this section to a proof that general Poisson brackets
(not assuming Hamilton's equations) satisfy the Jacobi identity. This is part of the thematic structure of this paper.
We are investigating the relationship of physics and its variables. When the variables are commutative it is a classical matter to 
have precise locations and standard coordinates. When the ``variables" are non-commutative one gives up the notion of location in
varying degrees, and gets the benefit of the extra mathematical structures in non-commutative worlds. The Poisson bracket is singular
in that it is a way to produce a Lie algebra structure from the algebra of derivations of a commutative algebra. Thus the Poisson
bracket is a key link betweem commutative and non-commutative worlds. This section is intended to make our story complete and to raise
the question of how this connection really comes about. Section 9 is a diagrammatic extension of section 8. We show how, in a
diagrammatic framework, the Jacobi identity can be articulated, and how it can arise from purely combinatorial grounds. These are
hints of further discrete physics. Section 10 is an epilogue, discussing the themes of the paper.
\bigbreak

\noindent {\bf Remark.} This paper is essentially self-contained, and hence it is written in an elementary
style. While there is a large literature on non-commutative geometry, emanating from the idea of replacing a space by its ring of 
functions, this paper is not written in that tradition. Non-commutative geometry does occur here, in the sense of geometry occuring
in the context of non-commutative algebra. Derivations are represented by commutators. There are relationships between the
present work and the traditional non-commutative geometry, but that is a subject for further exploration. In no way is this paper
intended to be an introduction to that subject.
\bigbreak

The following references in relation to non-commutative calculus are useful in 
comparing with our approach \cite{Connes, Dimakis, Forgy, MH}. Much of the present work is the fruit of a long
series of discussions with Pierre Noyes, and we will be preparing collaborative papers on it.  
I particularly thank Eddie Oshins for pointing out the relevance of minimal coupling. The
paper \cite{Mont} also works with minimal coupling for the Feynman-Dyson derivation. The first remark about
the minimal coupling occurs in the original paper by Dyson \cite{Dyson}, in the context of Poisson brackets.
The paper \cite{Hughes} is worth reading as a companion to Dyson. In the present paper we generalize the minimal
coupling to contexts including both commutators and Poisson brackets. It is the purpose of this paper to indicate how
non-commutative calculus can be used in foundations.
\bigbreak 

\noindent {\bf Acknowledgement.} Most of this effort was sponsored by the Defense
Advanced Research Projects Agency (DARPA) and Air Force Research Laboratory, Air
Force Materiel Command, USAF, under agreement F30602-01-2-05022.  The U.S. Government is authorized to reproduce and distribute
reprints for Government purposes notwithstanding any copyright annotations thereon. The
views and conclusions contained herein are those of the authors and should not be
interpreted as necessarily representing the official policies or endorsements,
either expressed or implied, of the Defense Advanced Research Projects Agency,
the Air Force Research Laboratory, or the U.S. Government. (Copyright 2004.) 
It gives the author great pleasure to acknowledge support from NSF Grant DMS-0245588,
and to give thanks to the University of Waterloo and the Perimeter Institute in Waterloo, Canada for their hospitality during
the preparation of this research.  
\bigbreak
  
\bigbreak

\section {Discrete Derivatives Become Commutators}
Consider a discrete deriviative $Df = (f(x+\Delta) - f(x))/\Delta.$
It is easy to see that $D$ does not satisfy the Leibniz rule. In fact, if
$$\tilde{f}(x) = f(x + \Delta),$$ then 
$$Df= (\tilde{f} - f)/\Delta$$ and one calculates that  
$$D(fg) = D(f)g + \tilde{f}D(g).$$ In the limit as $\Delta$ goes to zero,
$\tilde{f}$ approaches $f$ and the Leibniz rule is satisfied.
Now define a shift operator $J$ that satisfies the equation
$$Jf(x + \Delta) = f(x)J$$ or equivalently
$$J\tilde{f} = fJ.$$ Note that the existence of $J$ is accomplished by taking the commutative algebra $\cal C$ 
that we started with, and extending it to the free product of $\cal C$ with an algebra generated by the symbol $J,$ modulo
the ideal generated by  $fJ - J\tilde{f}$ for all $f$ in $\cal C.$
\bigbreak

\noindent Setting $$\nabla = JD,$$ we have
$$\nabla(fg) = JD(f)g + J\tilde{f}D(g) = JD(f)g + fJD((g).$$ Hence
$$\nabla(fg) = \nabla(f)g + f\nabla(g).$$ The adjusted derivative
$\nabla$ satisfies the Leibniz rule. 
\bigbreak

\noindent In fact, this adjusted derivative is a commutator in the algebra of
functions $\cal C$, extended by the operator J:
$$\nabla(f) = J(\tilde{f} - f)/\Delta = (fJ - Jf)/\Delta.$$ Hence 
$$\nabla(f) =[f,J/\Delta].$$  Note however that 
$$[x,J/\Delta] = (xJ  - Jx)/\Delta = J(x + \Delta - x)/\Delta = J.$$ Thus $\nabla(x) = J.$
This underlines the fact that these derivatives now take values in a non-commutative 
algebra. Note however, that if $$x^{(n)} = x(x - \Delta)...(x - (n-1)\Delta),$$
then $$\nabla(x^{(n)}) = JD((x^{(n)}) = Jnx^{(n-1)}.$$ Hence we can proceed in 
calculations with power series just as in ordinary discrete calculus, keeping in 
mind powers of $J$ that are shifted to the left. That is, a typical power series 
should be expressed in terms of the falling powers $x^{(n)}.$ We would define
$$exp_{\Delta}(x) = \Sigma_{n=0}^{\infty} x^{(n)}/n!$$ and find that 
$$\nabla(exp_{\Delta}(x)) = Jexp_{\Delta}(x).$$ The price paid for having the
Leibniz rule restored and the derivatives expressed in terms of commutators is the
appearance factors of $J$ on the left in final expressions of functions and
derivatives. 
\bigbreak

Note that we have $$\nabla(x) = [x,J/\Delta] = J,$$ and that this writes discrete 
calculus in terms that satisfy the Leibniz rule with a step of size $\Delta.$
It would be convenient to have an operator $P$ such that $[x,P] =1.$ Then
$[f,P]$ would formally mimic the usual derivative with respect to 
$x,$ and we would have $$[x^{n},P] = nx^{n-1}.$$ Of course, we can 
simply posit such a $P,$ but in fact, we can redefine $J$ so that
$$fJ = J\tilde{f}$$ where $$\tilde f(x) = f(x + J^{-1}\Delta).$$ Then
$$\nabla(x) = [x,J/\Delta] = J(x + J^{-1} \Delta - x)/\Delta  = 1$$ and we can take $$P = J/\Delta.$$ In this 
interpretation, $[f,P] = JDf = \nabla f$ where 
$$Df(x) = (f(x + J^{-1}\Delta) - f(x))/\Delta.$$
This double readjustment of the discrete derivative allows us to transfer 
standard calculus to an algebra of commutators. For physical applications however, there remains a difficulty in adding in a time
variable
$t$ and allowing that all the other elements of the algebra should be functions of time. If the derivative with respect to time is
represented by commutation with $H,$ then we cannot assume that $H$ commutes with $x.$ For this reason we will not proceed in the rest
of the  paper via this method of double readjustment.
\bigbreak

The cost for the double readjustment is that we must have a collection of functions in the original algebra $\cal C$ such that one
can sensibly define $\tilde f(x) = f(x + J^{-1}\Delta).$ Polynomial and power series functions have such natural extensions. For other
function algebras it will be an interesting problem in analysis, and algebra, to understand the structure of such extensions of 
commutative rings of functions to non-commutative rings of functions.
\bigbreak

\section{Time, Discrete Observation, Brownian Walks and the Simplest Commutator}
For temporal discrete derivatives there is a very neat interpretation of the shift operator of the previous section.
Consider a time series $\{X,X',X'',...\}$ with commuting scalar values.
Let $$DX = \dot{X} = J(X'-X)/\tau$$ where $\tau$ is an elementary time step (If $X$ denotes a times series value at time $t$, then 
$X'$ denotes the value of the series at time $t + \tau.$). The shift operator $J$ is defined by the equation
$XJ = JX'$
where this refers to any point in the time series so that $X^{(n)}J = JX^{(n+1)}$ for any non-negative integer $n.$
Moving $J$ across a variable from left to right, corresponds to one tick of the clock. We already know that this discrete,
non-commutative time derivative satisfies the Leibniz rule. 
\bigbreak

This derivative $D$ also fits a significant pattern of discrete observation. Consider the act of observing $X$ at a given time
and the act of observing (or obtaining) $DX$ at a given time. 
Since $X$ and $X'$ are ingredients in computing $(X'-X)/\tau,$ the numerical value associated with $DX,$ it is necessary  to let the
clock tick once, Thus, if we first observe
$X$ and then obtain $DX,$ the result is different (for the $X$ measurement) if we first obtain $DX,$ and then observe $X.$ In the
second case, we shall find the value $X'$ instead of the value $X,$ due to the tick of the clock. 
\bigbreak

\begin{enumerate}
\item Let $\dot{X}X$ denote the sequence: observe $X$, then obtain $\dot{X}.$ 
\item Let $X\dot{X}$ denote the sequence: obtain $\dot{X}$, then observe $X.$ 
\end{enumerate}
\bigbreak

\noindent We then see that the evaluation of these expressions in the
non-commutative calculus parallels the observational situation:
$$X\dot{X} = XJ(X'-X)/\tau = JX'(X'-X)/\tau$$
$$\dot{X}X = J(X'-X)X/\tau.$$
\bigbreak

The numerical evaluations for two such orderings are obtained by moving all occurrences of $J$ all the way to the left.
Thus we could write $$|J^{m}A| = A$$ for an expression where $A$ has no appearance of $J$. Then
$$|X\dot{X}| = X'(X'-X)/\tau$$ and 
$$|\dot{X}X| = (X'-X)X/\tau.$$
Elsewhere \cite{KN:QEM} we have called this interpretation of the temporal discrete derivative the ``Discrete Ordered Calculus"
or $DOC$ for short.
\bigbreak

The commutator $[X, \dot{X}]$ expresses the difference between these two orders of discrete measurement.
In the simplest case, where the elements of the time series are commuting scalars, we have
$$[X,\dot{X}] = X\dot{X} - \dot{X}X = XJ(X'-X)/\tau - J((X'-X)/\tau)X$$
$$= J[X'(X'-X) - (X'-X)X]/\tau =J(X'-X)^{2}/\tau.$$
Thus we can interpret the equation $$[X,\dot{X}] = Jk$$ ($k$ a constant scalar) as $$(X'-X)^{2}/\tau = k.$$ This means
that the process is a Brownian walk with spatial step $$\Delta = \pm \sqrt{k\tau}$$ where $k$ is a constant. In other words, we
have
$$k = \Delta^{2}/\tau.$$ 
We have shown that a Brownian walk with spatial step size  $\Delta$ and time step $\tau$ will satisfy the commutator equation above
exactly when the square of the spatial step divided by the time step remains constant. This means that {\em a given commutator equation
can be satisfied by walks with arbitrarily small spatial step and time step, just so long as these steps are in this fixed ratio.} 
\bigbreak

Remarkably, we can identify the constant $k/2$ as the {\it diffusion constant} for the Brownian process. 
To make this comparison, lets recall how the diffusion equation usually arises in discussing
Brownian motion. We are given a Brownian process where $$x(t + \tau) = x(t) \pm \Delta$$ 
\noindent so that the time step is $\tau$ and the space step is of absolute value $\Delta.$
We regard the probability of left or right steps as equal, so that if $P(x,t)$ denotes the
probability that the Brownian particle is at point $x$ at time $t$ then
$$P(x, t+\tau) = P(x-\Delta,t)/2 + P(x+\Delta, t)/2.$$ \noindent From this equation for the probability
we can write a difference equation for the partial derivative of the probability with respect
to time: 
$$(P(x, t + \tau) - P(x,t))/\tau = (h^{2}/2\tau)[(P(x-\Delta,t) - 2P(x,t) + P(x+\Delta))/\Delta^{2}] $$ 
\noindent The expression in brackets on the right hand side is a discrete approximation to
the second partial of $P(x,t)$ with respect to $x.$ Thus if the ratio $C = \Delta^{2}/2\tau$
remains constant as the space and time intervals approach zero, then this equation goes in 
the limit to the diffusion equation 
$$\partial P(x,t)/\partial t  = C \partial^{2}P(x,t)/\partial x^{2}.$$
$C$ is called the diffusion constant for the Brownian process.
\bigbreak

{\em The appearance of the diffusion constant from the observational commutator shows that this ratio is fundamental to the 
structure of the Brownian process itself, and not just to the probabilistic analysis of  that process.}
\bigbreak

\subsection{Planck's Numbers, Schr\"{o}dinger's Equation and the Diffusion Equation}
First recall the Planck Numbers. $\hbar$ is Planck's constant divided by $2\pi.$
$c$ is the speed of light.
$G$ is Newton's gravitational constant. The Planck length will be denoted by $L$, the Planck
time by $T$ and the Planck mass by $M.$ Their formulas are
$$M = \sqrt{\hbar c/G}$$
$$L = \hbar/Mc$$
$$T = \hbar/Mc^{2}.$$
\noindent These portions of mass, length and time are constructed
from the values of fundamental physical constants. They have roles in physics that point
to deeper reasons than the formal for introducing them. Here we shall see how they are related
to the Schr\"{o}dinger equation.
\bigbreak

Recall that Schr\"{o}dinger's equation can be regarded as the diffusion equation with an 
imaginary diffusion constant. Recall how this works. The Schr\"{o}dinger equation is
$$i\hbar \partial \psi/\partial t = H\psi$$ \noindent where the Hamiltonian $H$ is given by
the equation $H = p^{2}/2m + V$ where $V(x,t)$ is the potential energy and 
$p = (\hbar/i) \partial/\partial x$ is the momentum operator. With this we have
$p^{2}/2m =  (-\hbar^{2}/2m) \partial^{2}/\partial x^{2}.$ 
Thus with $V(x,t) = 0$, the equation becomes 
$i\hbar \partial \psi/\partial t = (-\hbar^{2}/2m) \partial^{2} \psi/\partial x^{2}$
which simplifies to
$$\partial \psi/\partial t = (i\hbar/2m) \partial^{2} \psi/\partial x^{2}.$$
\noindent Thus we have arrived at the form of the diffusion equation with an imaginary 
constant, and it is possible to make the identification with the diffusion equation by 
setting $$\hbar/m = \Delta^{2}/\tau$$ \noindent where $\Delta$ denotes a space interval, and
$\tau$ denotes a time interval as explained in the last section about the Brownian walk.
With this we can ask what space interval and time interval will satisfy this relationship? 
{\em Remarkably, the answer is that
this equation is satisfied when
$m$ is the Planck mass, $\Delta$ is the Planck length and $\tau$ is the Planck time.}
For note that $$L^{2}/T = (\hbar/Mc)^{2}/(\hbar/Mc^{2}) = \hbar/M.$$
\bigbreak

What does all this say about the nature of the Schr\"{o}dinger equation itself?
Interpreting it as a diffusion equation with imaginary constant suggests comparing with the
DOC equation $$[X,\dot{X}] = iJk$$ for a real constant $k$. This equation implicates a Brownian process
where $X'= X \pm \Gamma$ where $\Gamma^{2}/ \tau = ik.$ We can take $\Gamma = \sqrt{i}\Delta$ where $\Delta$ is a real
step-length. This gives a Brownian walk in the complex plane with the correct DOC diffusion 
constant. However, the relationship of this walk with the Schr\"{o}dinger equation is less 
clear because the $\psi$ in that equation is not the probability for the Brownian process.
To see a closer relationship we will take a different tack.
\bigbreak

Consider a discrete function $\psi(x,t)$ defined (recursively) by the following equation
$$\psi(x, t+\tau) = (i/2)\psi(x - \Delta,t) + (1-i)\psi(x,t) + (i/2)\psi(x + \Delta,t)$$
\noindent In other words, we are thinking here of a random ``quantum walk" where the amplitude
for stepping right or stepping left is proportional to $i$ while the amplitude for not moving
at all is proportional to $(1-i).$ It is then easy to see that $\psi$ is a discretization of 
$$\partial \psi/\partial t = (i\Delta^{2}/2\tau)\partial^{2} \psi/\partial x^{2}.$$
\noindent Just note that $\psi$ satisfies the difference equation
$$(\psi(x,t+\tau) - \psi(x,t))/\tau =
(i\Delta^{2}/2\tau)(\psi(x - \Delta,t) -2\psi(x,t) + \psi(x + \Delta,t))/\Delta^{2}$$
\noindent This gives a direct interpretation of the solution to the 
Schr\"{o}dinger equation as a limit of a sum
over generalized Brownian paths with complex amplitudes. We can then reinterpret this in 
DOC terms by the equation $[X,\dot{X}] = J(\Delta^{2}/\tau)$ or $[X,\dot{X}] = 0$, each of these
contingincies happening probabilistically. For a different (and deeper) relationship between Brownian motion and 
quantum mechanics see \cite{Nelson}. 
\bigbreak

\subsection{DOC Chaos}
Along with the simple Brownian motion solution to the one dimensional commutator equation, there is a 
heirarchy of time series that solve this equation, with periodic and chaotic behaviour. These solutions
can be obtained by taking 
$$X= J^{n}Y$$ where Y is a numerical scalar, and taking the commutator equation to be
$$[X, \dot{X}] = J^{2n+1}k$$ where $k$ is a scalar. Expanding this equation, we find
$$XJ(X'-X) - J(X'-X)X = J^{2n+1}k$$
$$J^{n}YJ(J^{n}Y' - J^{n}Y) - J(J^{n}Y' - J^{n}Y)J^{n}Y = J^{2n+1}k$$ 
$$J^{2n+1}Y^{n+1}(Y' - Y) - J^{2n+1}(Y^{n+1} - Y^{n})Y = J^{2n+1}k$$ 
$$Y^{n+1}(Y' - Y) - (Y^{n+1} - Y^{n})Y = k$$ 
$$Y^{n+1}(Y' - 2Y) = k - Y^{n}Y $$ 
$$Y^{n+1} = (k - Y^{n}Y)/(Y'-2Y).$$ This last equation expresses the time series recursively where $Y$ refers
to the value of the series that is $n$ time steps back from $Y^{n}.$ The first case of this recursion is
$$Y'' = (k - Y'Y)/(Y'-2Y).$$ Next case is
$$Y''' = (k - Y''Y)/(Y'-2Y).$$ These recursions depend critically on the value of the parameter $k.$ In the first
case one sees periodic oscillations that (for appropriate values of $k$) destabilize and blow up, alternating 
between an unbounded phase and a bounded semi-periodic phase. In Figures 1 and 2 we illustrate the case of the equation
$$Y'' = (k - Y'Y)/(Y'-2Y)$$ for $k=.0001$ in Figure 1 (a bounded phase) and $k=.009$ in Figure 2 (an unbounded phase).
There is intricate recursive structure in this hierarchy and it deserves further study.
\bigbreak

\begin{center}
$$ \picill4inby3in(Bounded)  $$
{ \bf Figure 1 -- $Y'' = (.0001 - Y'Y)/(Y'-2Y)$} \end{center}  
\bigbreak

\begin{center}
$$ \picill4inby3in(Unbounded)  $$
{ \bf Figure 2 -- $Y'' = (.009 - Y'Y)/(Y'-2Y)$} \end{center}  
\bigbreak

\section{Non Commutative Calculus and Hamilton's Equations}
We now set up a framework for non-commutative calculus in an arbitrary number of 
dimensions. We shall assume that each derivative is represented by a commutator, and that 
the basic space and time derivatives commute with one another as is customary for the flat space
of standard multi-variable calculus. This production of a flat space for calculus forms a reference domain
within the containing Lie algebra $\cal{A}.$  
\bigbreak

Since all derivatives are represented by commutators, this includes the time derivative as well.
We shall assume that there is an element $H$ in $\cal A$ representing the time derivative. This means
that $$dA/dt = [A,H]$$ for any $A$ in $\cal A$. Note that it follows at once from this choice that 
$H$ itself is time independent, since $dH/dt = [H,H] = 0.$ We shall see that $H$ behaves formally like 
the Hamiltonian operator in classical mechanics.
\bigbreak

We will assume that there is a set of  coordinates  $\{X_{1},...,X_{d}\}$ that are as flat as possible.
It is assumed that the $X_{i}$ all commute with one another, and that
the derivatives with respect to them commute with one another.
The partial derivatives with respect to $X_{i}$ will be represented by a set of elements
$\{ P_{1},\cdots, P_{d} \}$ with 
$$\partial_{i}F = \partial F/\partial X_{i} = [F,P_{i}]$$ for any $F$ in $\cal A.$
\bigbreak

\noindent Since we want the equation 
$$\partial X_{i}/\partial X_{j} = \delta_{ij},$$
we need the equation $$[X_{i},P_{j}] = \delta_{ij}.$$
\bigbreak

\noindent Since we want $$\partial_{i}\partial_{j} = \partial_{j}\partial_{i},$$
and since (as we compute in the introduction and in the next section)
$$\partial_{i}\partial_{j}F - \partial_{j}\partial_{i}F = [ \partial_{i}, \partial_{j} ]F = [ [ P_{i},P_{j} ] , F],$$
we see that these partial derivatives will commute with one another exactly when $[ P_{i},P_{j} ]$ belongs to the center of 
the algebra $\cal{A}$ for all choices of $i$ and $j.$ 
\bigbreak

\noindent For simplicity, we shall assume that 
$$[ P_{i},P_{j} ] = 0.$$
\bigbreak

\noindent With these choices the
flat coordinates satisfy:

$$[X_{i}, X_{j}] = 0$$
$$[P_{i},P_{j}]=0$$
$$[X_{i},P_{j}] = \delta_{ij}.$$
Note that we also have 
$$\hat{\partial_{i}}F = \partial F/\partial P_{i} = [X_{i},F]$$ so that
$$\hat{\partial_{j}}X_i = \partial X_{i}/\partial P_{j} = [X_{j}, X_{i}] = 0$$ and
$$\partial_{j} P_{i} = \partial P_{i}/\partial P_{j} = [X_{j}, P_{i}] =
\delta_{ij}.$$
\bigbreak

This formalism looks like bare quantum mechanics and can be so interpreted.
(if we take $i\hbar dA/dt = [A, H]$ and $H$ the Hamiltonian operator).
But these coordinates can also be viewed as the simplest flat set of
coordinates for referring the description of temporal phenomena in a non-commutative world.
There are various things to note. For example

$$dP_{i}/dt = [P_{i}, H] = -[H, P_{i}] = -\partial H/\partial X_{i}$$
$$dX_{i}/dt = [X_{i}, H] = \partial H/\partial P_{i}.$$ Thus
$$dP_{i}/dt = -\partial H/\partial X_{i}$$
$$dX_{i}/dt = \partial H/\partial P_{i}.$$

\noindent These are exactly Hamilton's equations of motion. The pattern of
Hamilton's equations is built into the system!
\bigbreak

\subsection{Hamilton's Equations in Classical Mechanics}
It is worth recalling how Hamilton's equations appear in classical mechanics. For simplicity, we shall restrict to one 
spatial variable $q$ (the analog of the operator $X$) and one momentum variable $p$ (the analog of the operator $P$).
In classical mechanics in one space and one time dimension, we have the equations
$$p = mdq/dt,$$
$$H = p^{2}/2m + V(q),$$
$$md^{2}q/dt^{2} = -\partial V/\partial q,$$
where the first equation is the definition of momentum of a particle of mass $m$, the second equation is the expression for the
energy of the system as the sum of the kinetic energy $p^{2}/2m$ and the potential energy $V(q).$ The third equation is Newton's 
law of motion.
\bigbreak

We see that $$\partial H/\partial p = p/m = dq/dt$$ and 
$$ \partial H /\partial q = \partial V/\partial q = - md^{2} q/dt^{2} = - dp/dt.$$
Thus
$$\partial H/\partial p = \dot{q}$$ and 
$$ \partial H /\partial q = - \dot{p}.$$ These are Hamilton's equations of motion.
\bigbreak

Hamilton went on to observe that for any function $F$ of $q$ and $p,$
$$\dot{F} = dF/dt = \partial F/\partial q \, \dot{q} + \partial F/\partial p \,  \dot{p}$$
$$= \partial F/\partial q \, \partial H/\partial p - \partial F/\partial p \, \partial H /\partial q.$$
Thus $$ \dot{F} = \{F,H \},$$ where $\{F,H \}$ is the Poisson Bracket defined by the equation
$$\{ A, B \} = \partial A/\partial q \, \partial B/\partial p - \partial A/\partial p \, \partial B /\partial q.$$
Remarkably, the Poisson Bracket satisfies the Jacobi identity, and hence gives a Lie algebra structure on the commutative space of 
functions of position and momentum. We have shown in this section that the pattern of Hamilton's equations is inherent in the LIe
algebra context. We shall have more to say about Poisson Brackets and Hamilton's equations in Section 5.
\bigbreak

\subsection{Curvature}
 \noindent  Note that for any $A,$ $B,$ $C$ in $\cal A$ we have the Jacobi Identity
$$[[A,B],C] + [[C,A],B] + [[B,C],A] = 0.$$  
\bigbreak

Suppose that $\{ \nabla_i \}$ is a collection of derivations on $\cal A$, represented
respectively by $\{ N_i \}$ so that $\nabla_i(F) = [F,N_i]$ for each $F$ in $\cal A.$
We define the {\em curvature} of the collection $\{ \nabla_i \}$ to be the collection
of commutators $\{ R_{ij} = [N_i,N_j] \}.$
\bigbreak      

\noindent {\bf Proposition 4.1} Let the family $\{ \nabla_i \}$ be given as above with
$\nabla_i(F) = [F,N_i].$ then
$$[\nabla_i, \nabla_j]F = [[N_i,N_j], F]$$ for all $F$ in $\cal A.$
Hence the curvature of $\{ \nabla_i \}$  measures the 
deviation of the cocatenations of these derivations from commutativity.
\bigbreak

\noindent {\bf Proof.} First, 
$$\nabla_i(\nabla_j(F)) = [[F,N_j],N_i],$$ which becomes via Jacobi identity
$$ = -[[N_j,N_i], F] - [[N_i,F], N_j]$$
$$ = [[N_i,N_j],F] + [[F,N_i], N_j].$$ Hence
$$ \nabla_i(\nabla_j(F)) =  [[N_i,N_j],F] + \nabla_j(\nabla_i(F)).$$ Whence
$$[\nabla_i, \nabla_j]F = [[N_i,N_j], F].$$
This proves the proposition. $\hfill\Box$
\bigbreak

This elementary notion of curvature for a collection of derivations just measures the extent to which they do not commute with 
one another. We have a collection $\cal{X}$ of elements $X$ of the algebra, and corresponding derivations
$\nabla_{X}$ where $\nabla_{X}Z = {Z, P_{X}}$ for a corresponding set of elements $P_{X}$ representing these derivations.
This is exactly the situation of the main framework of this section where
$$[X_{i}, X_{j}] = 0$$
$$[P_{i},P_{j}]=0$$
$$[X_{i},P_{j}] = \delta_{ij}$$
so that $$P_{X_{i}} = P_{i}.$$
In this case we have $\nabla_{i}Z = [Z, P_{i}],$ and the curvature of the collection is the collection of commutators
$\{ R_{ij} = [P_i,P_j] \}.$ More generally, the elements of the collection $\cal{X}$ may not commute with one another.
In this case we shall define the curvature as an operator $R(X,Y)$ defined on $\cal{X} \times \cal{X}$ by the equation
$$R(X,Y)Z = [\nabla_{X}, \nabla_{Y}]Z - \nabla_{[X,Y]}Z$$
in direct analogy with the usual definition in standard differential geometry. However, in order to do this we shall need
a collection $\cal{X}$ closed under the commutator, so that $\nabla_{[X,Y]}$ is defined. The next paragraphs outline one way to
accomplish this end.
\bigbreak

We will build the notion of curvature in terms of a general concept of covariant differentiation.
Let $\cal{X'}$ denote a specific collection of elements of the algebra $\cal{A}$ that we shall refer to as the
{\em variables} in $\cal{A}.$ We include in $\cal{A}$ a notion of {\em time} in the sense that there is
the temporal derivative $\dot{F} = [F,H]$ and we designate a single special variable $T$ to correspond to this temporal derivative.
In general, to each variable $X$ there is associated a derivative $\nabla_{X}$ with associated action
$$\nabla_{X}(F) = [F, P_{X}]$$ where $P_{X}$ represents this derivative. (Thus we take $P_{T} = H.$)
In this generality, we make no assumptions about the
commutativity of the variables or of the corresponding elements that represent the derivatives. 

We define the following set 
$\cal{S}$ of ``scalar functions over $\cal{X}$".
$${\cal S} = Center \{ f \in {\cal A} | [X,f] = 0 , X \in {\cal X'} \}.$$
That is, $\cal{S}$ is the set of $f$ that commute with each other and with all the elements of $\cal{X'}.$
 We then consider the closure of ${\cal X'}$ under addition and multiplication by elements
of ${\cal S}.$ Call this closure ${\cal X}.$ Define $P_{X + fY} = P_{X} + fP_{Y}$ for $X,Y \in {\cal X},
f \in {\cal S}.$
\bigbreak

For $f \in {\cal S}$ and $X \in {\cal X},$ define $X[f]$ by the formula
$$X[f] = \nabla_{X}(f) = [f, P_{X}].$$
Note that even though $f$ commutes with all the  elements of ${\cal X}$, it can still have non-trivial derivations
with respect to these variables.
\bigbreak

\noindent {\bf Lemma 4.2} We have the following properties:
\begin{enumerate}
\item For all $X,Y \in {\cal X}, \nabla_{X+Y} = \nabla_{X} + \nabla_{Y}.$
\item For all  $f \in {\cal S}, X,Y \in {\cal X}, \nabla_{fX}(Y) = f\nabla_{X}(Y).$
\item For all $f \in {\cal S}, X,Y \in {\cal X}, \nabla_{X}(fY) = X[f]Y + f\nabla_{X}(Y).$
\end{enumerate}
\bigbreak

\noindent {\bf Proof.} For the first property, note that 
$$\nabla_{X+Y}(Z) = [Z, P_{X+Y}] = [Z, P_{X} + P_{Y}] = \nabla_{X}(Z) + \nabla_{Y}(Z).$$
For the second property, we have
$$\nabla_{fX}(Y) = [Y, P_{fX}] = [Y,fP_{X}] = [Y,f]P_{X} + f[Y,P_{X}] = 0 + f\nabla_{X}(Y) = f\nabla_{X}(Y).$$
For the third property, note that 
$$\nabla_{X}(fY) = [fY, P_{X}] = [f,P_{X}]Y + f[Y, P_{X}] = X[f]Y + f\nabla_{X}(Y).$$
This completes the proof. $\hfill \Box$
\bigbreak

\noindent {\bf Definition.} We define the {\em curvature} as a function $R:{\cal X} \times {\cal X} \times {\cal X}
\longrightarrow {\cal A}$ by the formula $$R(X,Y)Z = [\nabla_{X}, \nabla_{Y}]Z - \nabla_{[X,Y]}Z.$$
Thus for given elements $X,Y \in {\cal X},$ the curvature operator $R(X,Y)$ measures the non-commutativity of the operators
$\nabla_{X}$ and $\nabla_{Y}$ in relation to the non-commutativity of $X$ and $Y.$ If $X$ and $Y$ commute, then 
$$R(X,Y)Z = [\nabla_{X}, \nabla_{Y}]Z ,$$ and we are returned to our initial definition of curvature for a collection of derivations.
\bigbreak

\section{General Equations of Motion}
Given a set of coordinates $\{X_1, X_2,\cdots X_d \}$ and dual coordinates $\{P_1, P_2,\cdots P_d \}$  as in the previous section,
a general description of $dX_{i}/dt$ takes the form of a system of  equations
$$dX_{i}/dt = {\cal G}_{i}$$ where $\{ {\cal G}_{1},\cdots, {\cal G}_{d} \}$
is a collection of elements of $\cal A.$ If we choose to write ${\cal G}_{i}$
relative to the flat coordinates via ${\cal G}_{i} = P_{i} -  A_{i}$
(this is a definition of $A_{i}$) then the formalism of gauge theory appears
naturally.
For example, if $$\nabla_{i}(F) = [F, {\cal G}_{i}],$$ then we have
the curvature $$[\nabla_{i}, \nabla_{j}]F = [R_{ij}, F]$$
where
$$R_{ij} = [{\cal G}_{i}, {\cal G}_{j}]$$ 
$$= [P_{i} -  A_{i}, P_{j} -  A_{j}]$$
$$= -[P_i, A_j] -[A_i,P_j] + [A_i, A_j]$$
$$= \partial_{i} A_{j} - \partial_{j} A_{i} + [A_{i}, A_{j}].$$  This is the
well-known formula that expresses the gauge field as the curvature of the gauge 
connection. From this point of view everything comes naturally from the assumption
that all derivatives are represented by commutators, and that one refers
all equations to the flat background coordinates.
\bigbreak

\subsection{Curvature and Connection at the Next Level}

The dynamical law is $$dX_{i}/dt = \dot{X_{i}} = P_i - A_i = {\cal G}_i.$$ This gives rise
to new  commutation relations 
$$[X_i, \dot{X_j}] = [X_i, P_j] - [X_i, A_j] = \delta_{ij} - \partial A_{j}/\partial P_{i} = g_{ij}$$ 
where this equation defines $g_{ij}$, and
$$[\dot{X_i}, \dot{X_j}] =R_{ij} =  \partial_{i} A_{j} - \partial_{j} A_{i} +
[A_{i}, A_{j}] .$$
\bigbreak

\noindent We define the ``covariant derivative" $$\nabla_{i}F = [F,P_i - A_i]
= \partial_{i}(F) - [F,A_i] = [F, \dot{X_i}],$$ while we can still write
$$\hat{\partial_i}F = [X_i, F].$$
\bigbreak

It is natural to think that $g_{ij}$ is analogous to a metric. This analogy is 
strongest if we {\em assume} that $$[X_i, g_{jk}] = 0.$$ By assuming that 
the spatial coordinates commmute with the metric coefficients we have that
$$[\dot{X_i}, g_{jk}] + [X_i, \dot{g_{jk}}] = 0.$$ Hence
$$\nabla_{i}g_{jk} = \hat{\partial_i}\dot{g_{jk}}.$$
Here, we shall let 
$$c_{ijk} = [X_i, g_{jk}],$$
not assuming this commutator to vanish. Then
$$\dot{c_{ijl}} = [\dot{X_i}, g_{jk}] + [X_i, \dot{g_{jk}}] = \hat{\partial_i}\dot{g_{jk}} - \nabla_{i}g_{jk}.$$
\bigbreak

A stream of consequences then follows by differentiating both sides of
the equation 

$$g_{ij}= [X_{i}, \dot{X_{j}}].$$ 
We will detail these consequences in section 6. For now, we show how the form of the Levi-Civita connection
appears naturally in this context.

\noindent In the following we shall use $D$ as an abbreviation for $d/dt.$
\bigbreak

The Levi-Civita connection 

$$\Gamma_{ijk} =(1/2)(\nabla_{i}g_{jk}
+\nabla_{j}g_{ik}-\nabla_{k}g_{ij})$$  

\noindent associated with the $g_{ij}$ comes up almost at once from the
differentiation process described above.  To see how this happens, view
the following calculation where 
$$\hat{\partial_{i}}\hat{\partial_{j}}F = [X_{i}, [ X_{j}, F]].$$
\bigbreak

\noindent We apply the operator $\hat{\partial_{i}}\hat{\partial_{j}}$ to the second
time derivative of
$X_{k}.$
\bigbreak

\noindent {\bf Lemma 5.1} Let $ \Gamma_{ijk} =(1/2)(\nabla_{i}g_{jk}
+\nabla_{j}g_{ik}-\nabla_{k}g_{ij}).$ Then
$$\Gamma_{ijk}=
(1/2)(\hat{\partial_{i}}\hat{\partial_{j}}\ddot{X_{k}} - \hat{\partial_{i}}g_{jk}),$$
where $c_{ijk} = [X_i, g_{jk}] = \hat{\partial_{i}}g_{jk}.$   
\bigbreak

\noindent {\bf Proof.} 
\noindent Note that by the Leibniz rule

$$D([A,B]) = [\dot{A}, B] + [A, \dot{B}],$$ we have

$$\dot{g_{jk}}= [\dot{X_{j}}, \dot{X_{k}}] + [X_{j}, \ddot{X_{k}}].$$
Therefore
$$\hat{\partial_{i}}\hat{\partial_{j}}\ddot{X_{k}} = [X_{i}, [ X_{j}, \ddot{X_{k}}]]$$

$$= [X_{i}, \dot{g_{jk}} - [\dot{X_{j}}, \dot{X_{k}}]]$$

$$= [X_{i}, \dot{g_{jk}}] -  [X_{i}, [\dot{X_{j}}, \dot{X_{k}}]]$$

$$= [X_{i}, \dot{g_{jk}}] + [\dot{X_{k}}, [X_{i}, \dot{X_{j}}]]+
[\dot{X_{j}}, [ \dot{X_{k}}, X_{i}]$$

$$= \dot{c_{ijl}} - [\dot{X_i}, g_{jk}] + [\dot{X_{k}}, [X_{i}, \dot{X_{j}}]]+
[\dot{X_{j}}, [ \dot{X_{k}}, X_{i}]$$

$$= c_{ijk} + \nabla_{i}g_{jk} -\nabla_{k}g_{ij} + \nabla_{j}g_{ik}$$

$$ = c_{ijk} + 2\Gamma_{kij}.$$
This completes the proof. $\hfill \Box$
\bigbreak

\noindent It is remarkable that the form of the Levi-Civita connection
comes up directly from this non-commutative calculus without any apriori
geometric interpretation. 
\bigbreak

The upshot of this derivation is that it confirms our interpretation of 
$$g_{ij} = [X_i, \dot{X_j}] = [X_i, P_j] - [X_i, A_j] = \delta_{ij} - \partial A_{j}/\partial P_{i}$$
as an abstract form of metric (in the absence of any actual notion of distance in the non-commutative 
world). {\em This calls for a re-evaluation and reconstruction of differential geometry based on non-commutativity and
the Jacobi identity.} This is differential geometry where the fundamental concept is no longer parallel translation, but 
rather a non-commutative version of a physical trajectory. This approach will be
the subject of a separate paper.
\bigbreak

At this stage we face the mystery of the appearance of the Levi-Civita connection.
There is a way to see that the appearance of this connection is not an accident, but
rather quite natural. We shall explain this point of view in the next subsection where we discuss
Poisson brackets and the connection of this formalism with classical physics. On the other hand,
we have seen in this section that it is quite natural for curvature in the form of the 
non-commutativity of derivations to appear at the outset in a non-commutative formalism. We have
also seen that this curvature and connection can be understood as a measurement of the deviation 
of the theory from the ``flat" commutation relations of ordinary quantum mechanics. 
Electromagnetism and Yang-Mills theory can be seen as the theory of the curvature introduced by 
such a deviation. On the other hand, from the point of view of metric differential geometry, the
Levi-Civita connection is the unique connection that preserves the inner product defined by the 
metric under the parallel translation defined by the connection. We would like to see that the
formal Levi-Civita connection produced here has this property as well.
\bigbreak

To this end lets recall the formalism of parallel translation. The infinitesimal parallel
translate of $A$ is denoted by $A' = A + \delta A$ where 
$$\delta A^{k} = -\Gamma^{k}_{ij}A^{i}dX^{j}$$ \noindent where here we are writing in the usual
language of vectors and differentials with the Einstein summation convention for repeated
indices. We assume that the Christoffel symbols satisfy the symmetry condition
$\Gamma^{k}_{ij} = \Gamma^{k}_{ji}.$ The inner product is given by the formula
$$<A,B> = g_{ij}A^{i}B^{j}$$ \noindent Note that here the bare symbols denote vectors whose 
coordinates may be indicated by indices. The requirement that this inner product be invariant
under parallel displacement is the requirement that $\delta(g_{ij}A^{i}A^{j}) = 0.$
Calculating, one finds 
$$\delta(g_{ij}A^{i}A^{j}) = (\partial_{k}g_{ij})A^{i}A^{j}dX^{k} + g_{ij}\delta(A^{i})A^{j} 
+ g_{ij}A^{i}\delta(A^{j})$$

$$= (\partial_{k}g_{ij})A^{i}A^{j}dX^{k} - g_{ij}\Gamma^{i}_{rs}A^{r}dX^{s}A^{j} 
- g_{ij}A^{i}\Gamma^{j}_{rs}A^{r}dX^{s}$$

$$= (\partial_{k}g_{ij})A^{i}A^{j}dX^{k} - g_{ij}\Gamma^{i}_{rs}A^{r}A^{j}dX^{s} 
- g_{ij}\Gamma^{j}_{rs}A^{i}A^{r}dX^{s}$$

$$= (\partial_{k}g_{ij})A^{i}A^{j}dX^{k} - g_{sj}\Gamma^{s}_{ik}A^{i}A^{j}dX^{k} 
- g_{is}\Gamma^{s}_{jk}A^{i}A^{j}dX^{k}$$

\noindent Hence

$$(\partial_{k}g_{ij}) = g_{sj}\Gamma^{s}_{ik} + g_{is}\Gamma^{s}_{jk}.$$

\noindent From this it follows that 

$$\Gamma_{ijk} = g_{is}\Gamma^{s}_{jk} = (1/2)(\partial_{k}g_{ij} - \partial_{i}g_{jk} 
+\partial_{j}(g_{ik})).$$

\noindent Certainly these notions of variation can be imported into our abstract context.
The question remains how to interpret the new connection that arises. We now have a new 
covariant derivative in the form 

$$\hat{\nabla_{i}}X^{j} = \partial_{i}X^{j} + \Gamma^{j}_{ki}X^{k}.$$

\noindent The question is how the curvature of this connection interfaces with the gauge 
potentials that gave rise to the metric in the first place. The theme of this investigation has
the flavor of gravity theories with a qauge theoretic background. We will investigate these 
relationships in a sequel to this paper.
\bigbreak

\subsection{Poisson Brackets and Commutator Brackets} 

Dirac \cite{Dirac} introduced a fundamental relationship between quantum
mechanics and classical mechanics that is summarized by the maxim {\em
replace Poisson brackets by commutator brackets.} Recall that the Poisson
bracket $\{ A, B\}$ is defined by the formula 

$$\{ A, B \} = (\partial A/ \partial q) (\partial B/ \partial p) -
(\partial A/ \partial p) (\partial B/ \partial q),$$ 

\noindent where $q$ and $p$ denote classical position and momentum
variables respectively.
\vspace{3mm}

In our version of discrete physics the noncommuting variables are
functions of time, with the time derivative itself a commutator. With $$DF=[F,J/\tau],$$ it
follows that $$D([A,B]) = [DA,B] + [A, DB]$$ for any expressions $A$, $B$
in our Lie Algebra. A corresponding Leibniz rule for Poisson brackets would
read
$$(d/dt) \{ A, B \} = \{ dA/dt , B \} + \{ A, dB/dt \}.$$ 
\vspace{3mm}

\noindent However, here there is an easily verified exact formula:
$$(d/dt) \{ A, B \} = \{ dA/dt , B \} + \{ A, dB/dt \} - \{ A, B
\}(\partial \dot{q} / \partial q + \partial \dot{p} / \partial p).$$ 

\noindent This means that the Leibniz formula will hold for the Poisson
bracket exactly when
$$(\partial \dot{q} / \partial q + \partial \dot{p} / \partial p)=0.$$ 
\vspace{3mm}

\noindent This is an integrability condition that will be satisfied if
$p$ and $q$ satisfy Hamilton's equations
$$ \dot{q} = \partial H / \partial p ,$$ $$ \dot{p} = - \partial H /
\partial q. $$ 
\vspace{3mm}

\noindent This means that $q$ and $p$ are following a
principle of least action with respect to the Hamiltonian $H$. Thus we
can interpret the {\em fact} $D([A,B]) = [DA,B] + [A, DB]$ in the
non-commutative context as an analog of the principle of least action. Taking
the non-commutative context as fundamental, we say that Hamilton's equations are
{\em motivated} by the presence of the Leibniz rule for the discrete
derivative of a commutator. The classical laws are obtained by following
Dirac's maxim in the opposite direction! Classical physics is produced by
following the correspondence principle upwards from the discrete.
\vspace{3mm}

In making this backwards journey to classical physics we see how our earlier assertion
that bare mechanics of commutators can be regarded as the background for the
coupling with other fields (as in the description of formal gauge theory),
fits with Poisson brackets. The bare Poisson brackets satisfy
$$\{q_{i}, q_{j}\}=0$$
$$\{p_{i}, p_{j}\}=0$$
$$\{q_{i}, p_{j}\} = \delta_{ij}.$$ In our previous formalism, we would identify
$X_{i}$ as the correspondent with $q_{i}$  and $P_j$ as the
correspondent of $p_{j}.$ And, given a classical vector potential $A$, we could write the 
coupling $dq_{i}/dt = p_{i} - A_i$ to describe the motion of a particle in the presence of an 
electromagnetic field.  Similar remarks apply to the analogs for gauge theory and curvature.
In particular it is of interest to see that our derivation of the Levi-Civita connection 
corresponds to the motion of a particle in generalized coordinates that satisfies Hamilton's equations.
The fact that such a particle moves in a geodesic according to the Levi-Civita connection is a classical
fact. Our derivation of the Levi-Civita connection, interpreted in Poisson brackets, reproduces this result.
\bigbreak

To see how this works, let $ds^2 = g^{ij}dx_i dx_j$ denote the metric in the generalized coordinates
$x_k.$ Then the velocity of the particle has square $v^2 = (ds/dt)^2 = g^{ij}\dot{x_i}\dot{x_j}.$
The Lagrangian for the system is the kinetic energy $L = mv^{2}/2 = mg^{ij}\dot{x_i}\dot{x_j}/2.$
Then the canonical
momentum is $p_j = \partial L/ \partial\dot{x_j},$ and with $q_i = x_i$ we have the Poisson brakets
$$\delta_{ij} = \{q_i, p_j\} = \{x_i, \partial L/ \partial\dot{x_j}\} = \{x_i, mg^{jk}\dot{x_k}\}.$$ Taking $m=1$
for simplicity, we can rewrite this bracket as $$\{x_i,\dot{x_j}\} = g_{ij}.$$ This, in Poisson brackets, is
our generalized equation of motion. \bigbreak

The classical derivation applies Lagrange's equation of motion to the system. Lagrange's equation reads
$$d/dt(\partial L/\partial \dot{x_i}) = \partial L/\partial x_i.$$ Since this equation is equivalent to 
Hamilton's equation of motion, it follows that the Poisson brackets satisfy the Leibniz rule. With this, we
can proceed with our derivation of the Levi-Civita connection in relation to the acceleration of the particle.
In the classical derivation, one writes out the Lagrange equation and solves for the acceleration.
The advantage of using only the Poisson brackets is that it shows the relationship of the connection with the
Jacobi identity and the Leibniz rule. \bigbreak

This discussion raises further questions about the
nature of the generalization that we have made. Originally Hermann Weyl
\cite{Weyl}  generalized classical differential geometry and discovered
gauge theory by allowing changes of length as well as changes of angle to
appear in the holonomy. Here we arrive at a similar situation via
the properties of a non-commutative discrete calculus of observations. 
\vspace{3mm}

\section{Consequences of the Metric}
In this section we shall follow the formalism of the metric commutator equation
$$ [X_{i}, \dot{X_{j}}] = g_{ij}$$
very far in a semi-classical context. That is, we shall set up a non-commutative world, 
{\em and} we shall make assumptions about the non-commutativity that bring the operators into close analogy with variables in standard 
calculus. In particular we shall regard an element $F$ of the Lie algebra to be a ``function of the $X_{i}$ " if $F$ commutes with
the $X_{i},$ and we shall assume that if $F$ and $G$ commute with the $X_{i},$ then $F$ and $G$ commute with each other. We call
this the {\em principle of commutativity}. With these background assumptions, it is possible to get a very sharp result about the 
behaviour of the theory. In particular, the results of this section sharpen the work in \cite{Tanimura}, where special orderings and
averages of orderings of the operators were needed to obtain analogous results.
\bigbreak

We assume that 
$$[X_{i}, \dot{X_{j}}] = g_{ij}$$
$$[X_i,X_j] = 0$$
$$[X_i,g_{jk}]=0$$ 
$$[g_{ij}, g_{kl}] = 0.$$
We assume that there exists a $g^{ij}$ with $$g^{ij}g_{jk} = \delta_{k}^{i} = g_{ij}g^{jk} = \delta^{k}_{i}.$$
We also assume that if $$[A,X_i] = 0$$ and $$[B , X_i] = 0$$ for all $i,$ then $$[A,B]=0$$ for all expressions $A$ and $B$
in the algebra under consideration. To say that $[A,X_i] = 0$ is to say the analogue of the statement that $A$ is a function
only of the variables $X_{i}$ and not a function of the $\dot{X_{j}}.$ This is a stong assumption about the algebraic 
structure, and it will not be taken when we look at strictly discrete models. It is, however, exactly the assumption that brings the
non-commutative algebra closest to the classical case of functions of positions and momenta.
\bigbreak

The main result of this section will be a proof that  
$$\ddot{X_{r}} = G_{r} + F_{rs}\dot{X^{s}} + \Gamma_{rst}\dot{X^{s}}\dot{X^{t}},$$
and that this decompositon of the acceleration is uniquely determined by the given framework. Since 
$$F^{rs} = [\dot{X^{r}}, \dot{X^{s}}]=g^{ri}g^{sj}F_{ij},$$ we can regard this result as a description of the motion of the 
non-commutative particle influenced by a scalar field $G_{r},$ a qauge field $F^{rs},$ and geodesic motion with respect to the 
Levi-Civita connection corresponding to $g_{ij}.$  Let us begin.
\bigbreak

\noindent Note that, as before, we have that $g_{ij} = g_{ji}$ by taking the time derivative of the equation 
$[X_i,X_j] = 0.$
\bigbreak

\noindent Note also that the Einstein summation convention (summing over repeated indices) is in effect when we write equations,
unless otherwise specified.
\bigbreak

As before, we define $$\partial_{i}F = [F, \dot{X_{i}}]$$ and $$\hat{\partial_{i}}F = [X_{i},F].$$
We also make the definitions $$\dot{X^{i}} = g^{ij}\dot{X_{j}}$$ and $$\partial^{i}F = [F,\dot{X^{i}}].$$
Note that we do not assume the existence of a variable $X^{j}$ whose time derivative is $\dot{X^{j}}.$
Note that we have $$\dot{X_{k}} = g_{ki}\dot{X^{i}}.$$
Note that it follows at once that $$\hat{\partial_{i}}\dot{g_{jk}} = \partial_{i}g_{jk}$$ by differentiating the equation
$[X_i,g_{jk}]=0.$ 
\bigbreak

We assume the following postulate about the time derivative of an element $F$ with $[X_{i}, F]=0$ for all $k:$
$$\dot{F} = (\partial_{i}F)\dot{X^{i}}.$$ This is in accord with the concept that $F$ is a function of the variables
$X_{i}.$ Note that in one interpretation of this formalism, one of the variables $X_{i}$ could be itself a time variable.
In the next section, we shall return to three dimensions of space and one dimension of time, with a separate notation for the time
variable. Here there is no restriction on the number of independent variables $X_{i}.$
\bigbreak

We have the following Lemma. 
\bigbreak

\noindent {\bf Lemma 6.1} \begin{enumerate}
\item $[X_{i},\dot{ X^{j}}] = \delta_{i}^{j}.$
\item $\partial_{r}(g^{ij})g_{jk} + g^{ij}\partial_{r}(g_{jk}) = 0.$
\item $[X_{r}, \partial_{i}g_{jk}] = 0.$
\end{enumerate}

\bigbreak

\noindent {\bf Proof.} $$[X_{i},\dot{ X^{j}}] = [X_{i},g^{jk}\dot{X_{k}}] = [X_{i},g^{jk}]\dot{X_{k}} + g^{jk}[X_{i},\dot{X_{k}}]$$
$$= g^{jk}[X_{i},\dot{X_{k}}] = g^{jk}g_{ik} = g^{jk}g_{ki} = \delta^{j}_{i}.$$  The second part of the proposition
is an application of the Leibniz rule:
$$0 = \partial_{r}(\delta^{i}_{k}) = \partial_{r}(g^{ij}g_{jk}) = \partial_{r}(g^{ij})g_{jk} + g^{ij}\partial_{r}(g_{jk}).$$
Finally, $$[X_{r}, \partial_{i}g_{jk}] = [X_{r}, [g_{jk}, \dot{X_{i}}]] = 
-[\dot{X_{i}}, [X_{r}, g_{jk}]] - [g_{jk}, [\dot{X_{i}}, X_{r}]]$$
$$ = -[\dot{X_{i}}, 0] + [g_{jk}, g_{ir}] = 0 + 0 = 0.$$
This completes the proof of the Lemma.
$\hfill \Box$
\bigbreak

\noindent It follows from this lemma that $\partial^{i}$ can be regarded as $\partial/\partial X_{i}.$
\bigbreak

We have seen that it is natural to consider the commutator of the velocities $R_{ij}=[\dot{X_{i}}, \dot{X_{j}}]$ as a field or
curvature. For the present analysis, we would prefer the field to commute with all the variables $X_k$ in order to identify it as a 
``function of the variables $X_{k}".$ We shall find, by a computation, that $R_{ij}$ does not so commute, but that a compensating 
factor arises naturally. The result is as follows.
\bigbreak

\noindent {\bf Proposition 6.2} Let $F_{rs} = [\dot{X_{r}}, \dot{X_{s}}] +(\partial_{r}g_{ks} - \partial_{s}g_{kr})X^{k}$
and $F^{rs} = [\dot{X^{r}}, \dot{X^{s}}].$ Then
\begin{enumerate}
\item $F_{rs}$ and $F^{rs}$ commute with the variables $X_{k}.$
\item $F^{rs} = g^{ri}g^{sj}F_{ij}.$
\end{enumerate}
\bigbreak

\noindent {\bf Proof of Proposition.}

We begin by computing the commutator of $X_{i}$ and $R_{rs} = [\dot{X_{r}}, \dot{X_{s}}]$ by using the Jacobi identity.
$$[X_{i}, [\dot{X_{r}}, \dot{X_{s}}]] = -[\dot{X_{s}}, [X_{i}, \dot{X_{r}}]] - [\dot{X_{r}}, [\dot{X_{s}}, X_{i}]]
= \partial_{s}g_{ir} -\partial_{r}g_{is}.$$ 
Note also that $$[X_{i}, \partial_{r}g_{ks}] = [X_{i}, [g_{ks}, \dot{X_{r}}]] =
-[\dot{X_{r}},[X_{i},g_{ks}]] -[g_{ks},[\dot{X_{r}}, X_{i}]]$$ 
$$= -[\dot{X_{r}},[X_{i},g_{ks}]] + [g_{ks},g_{ir}] = 0.$$
Hence $$[X_{i}, (\partial_{r}g_{ks} - \partial_{s}g_{kr})X^{k}] = \partial_{r}g_{is} - \partial_{s}g_{ir}.$$ 
Therefore $$[X_{i}, F_{rs}] = [X_{i}, [\dot{X_{r}}, \dot{X_{s}}] + (\partial_{r}g_{ks} - \partial_{s}g_{kr})X^{k} ] = 0.$$
This, and an a similar computation that we leave for the reader, proves the first part of the proposition.
We prove the second part by direct computation:
Note the following identity:
$$[AB,CD] = [A,C]BD + A[B,C]D + C[A,D]B + CA[B,D].$$ Using this identity we find
$$[\dot{X^{r}}, \dot{X^{s}}] = [g^{ri}\dot{X_{i}}, g^{sj}\dot{X_{j}}]$$
$$ = [g^{ri},g^{sj}]\dot{X_{i}}\dot{X_{j}} + g^{ri}[\dot{X_{i}},g^{sj}]\dot{X_{j}} + g^{sj}[g^{ri},\dot{X_{j}}]\dot{X_{i}} +
g^{sj}g^{ri}[\dot{X_{i}},\dot{X_{j}}]$$
$$= -g^{ri}\partial_{i}(g^{sj})\dot{X_{j}} + g^{sj}\partial_{j}(g^{ri})\dot{X_{i}} + g^{sj}g^{ri}[\dot{X_{i}},\dot{X_{j}}]$$
$$= -g^{ri}\partial_{i}(g^{sj})g_{jl}\dot{X^{l}} + g^{sj}\partial_{j}(g^{ri})g_{il}\dot{X^{l}} +
g^{sj}g^{ri}[\dot{X_{i}},\dot{X_{j}}]$$
$$= g^{ri}g^{sj}\partial_{i}(g_{jl})\dot{X^{l}} - g^{sj}g^{ri}\partial_{j}(g_{il})\dot{X^{l}} +
g^{sj}g^{ri}[\dot{X_{i}},\dot{X_{j}}]$$
$$= g^{ri}g^{sj}(\partial_{i}(g_{jl})\dot{X^{l}} - \partial_{j}(g_{il})\dot{X^{l}} + [\dot{X_{i}},\dot{X_{j}}])$$
$$= g^{ri}g^{sj}F_{ij}.$$ This completes the proof of the proposition. $\hfill \Box$
\bigbreak

We now consider the full form of the acceleration terms $\ddot{X_{k}}.$ We have already shown that 
$$\hat{\partial_{i}}\hat{\partial_{j}}\ddot{X_{k}} = \partial_{i}g_{jk}+\partial_{j}g_{ik}-\partial_{k}g_{ij}.$$ 
Letting $$\Gamma_{kij} = (1/2)(\partial_{i}g_{jk}+\partial_{j}g_{ik}-\partial_{k}g_{ij}),$$
we {\em define} $G_{r}$ by the formula
$$\ddot{X_{r}} = G_{r} + F_{rs}\dot{X^{s}} + \Gamma_{rst}\dot{X^{s}}\dot{X^{t}}.$$
\bigbreak

\noindent {\bf Proposition 6.3} Let $\Gamma_{rst}$ and $G_{r}$ be defined as above. Then both $\Gamma_{rst}$ and $G_{r}$
commute with the variables $X_{i}.$
\bigbreak

\noindent {\bf Proof.}  Since we know that $[X_{l}, \partial_{i}g_{jk}] = 0,$ it follows at once that
$[X_{l}, \Gamma_{rst}] = 0.$ It remains to examine the commutator $[X_{l}, G_{r}].$ We have
$$[X_{l}, G_{r}] = [X_{l},\ddot{X_{r}} - F_{rs}\dot{X^{s}} - \Gamma_{rst}\dot{X^{s}}\dot{X^{t}}]$$ 
$$= [X_{l},\ddot{X_{r}}] - [X_{l},F_{rs}\dot{X^{s}}] - [X_{l},\Gamma_{rst}\dot{X^{s}}\dot{X^{t}}]$$
$$= [X_{l},\ddot{X_{r}}] - F_{rs}[X_{l},\dot{X^{s}}] - \Gamma_{rst}[X_{l},\dot{X^{s}}\dot{X^{t}}]$$
(since $F_{rs}$ and $\Gamma_{rst}$ commute with $X_{l}$). Note that 
$$[X_{l},\dot{X^{s}}] = \delta^{s}_{l}$$ and that
$$[X_{l},\dot{X^{s}}\dot{X^{t}}]= [X_{l},\dot{X^{s}}]\dot{X^{t}} + \dot{X^{s}}[X_{l},\dot{X^{t}}]$$
$$= \delta^{s}_{l}\dot{X^{t}} + \dot{X^{s}}\delta^{t}_{l}.$$ Thus
$$[X_{l}, G_{r}] = [X_{l},\ddot{X_{r}}] - F_{rs}\delta^{s}_{l} - \Gamma_{rst}(\delta^{s}_{l}\dot{X^{t}} +
\dot{X^{s}}\delta^{t}_{l})$$
$$= [X_{l},\ddot{X_{r}}] - F_{rl} - \Gamma_{rlt}\dot{X^{t}} - \Gamma_{rsl}\dot{X^{s}}.$$
It is easy to see that $\Gamma_{rlt}\dot{X^{t}} = \Gamma_{rsl}\dot{X^{s}}.$ Hence
$$[X_{l}, G_{r}] = [X_{l},\ddot{X_{r}}] - F_{rl} - 2\Gamma_{rlt}\dot{X^{t}}.$$
On the other hand,
$$[X_{l}, \dot{X_{r}}] = g_{lr}.$$ Hence
$$[X_{l}, \ddot{X_{r}}] = \dot{g_{lr}} - [\dot{X_{l}}, \dot{X_{r}}].$$
Therefore
$$[X_{l}, G_{r}] = \dot{g_{lr}} - [\dot{X_{l}}, \dot{X_{r}}] - F_{rl} - 2\Gamma_{rlt}\dot{X^{t}}$$
$$= \dot{g_{lr}} - (\partial_{r}g_{kl} - \partial_{l}g_{kr})\dot{X^{k}} - 2\Gamma_{rlt}\dot{X^{t}}.$$
(since $F_{rl} = [\dot{X_{r}}, \dot{X_{l}}] +(\partial_{r}g_{kl} - \partial_{l}g_{kr})\dot{X^{k}}$)
Hence
$$[X_{l}, G_{r}] = \dot{g_{lr}} - (\partial_{r}g_{tl} - \partial_{l}g_{tr})\dot{X^{t}} -
(\partial_{l}g_{tr}+\partial_{t}g_{lr}-\partial_{r}g_{lt})\dot{X^{t}}$$
$$= \dot{g_{lr}} - (\partial_{t}g_{lr})\dot{X^{t}} = 0.$$
This completes the proof of the proposition. $\hfill \Box$
\bigbreak

We now know that $G_{r},$ $F_{rs}$ and $\Gamma_{rst}$ commute with the variables $X_{k}.$ As we now shall see,
the formula
$$\ddot{X_{r}} = G_{r} + F_{rs}\dot{X^{s}} + \Gamma_{rst}\dot{X^{s}}\dot{X^{t}}$$ allows us to extract these functions
from $\ddot{X_{r}}$ by differentiating with respect to the dual variables. We already know that 
$$\hat{\partial_{i}}\hat{\partial_{j}}\ddot{X_{k}} = 2\Gamma_{kij},$$ and now note that
$$\hat{\partial_{i}}(\ddot{X_{r}}) = [X_{i},\ddot{X_{r}}] = [X_{i},G_{r} + F_{rs}\dot{X^{s}} + \Gamma_{rst}\dot{X^{s}}\dot{X^{t}}]$$ 
$$= F_{rs}[X_{i},\dot{X^{s}}] + \Gamma_{rst}[X_{i},\dot{X^{s}}\dot{X^{t}}]$$
$$= F_{ri} + 2\Gamma_{rit}\dot{X^{t}}.$$
\bigbreak

We see now that the decomposition $$\ddot{X_{r}} = G_{r} + F_{rs}\dot{X^{s}} + \Gamma_{rst}\dot{X^{s}}\dot{X^{t}}$$
of the acceleration is uniquely determined by these conditions. Since 
$$F^{rs} = [\dot{X^{r}}, \dot{X^{s}}]=g^{ri}g^{sj}F_{ij},$$ we can regard this result as a description of the motion of the 
non-commutative particle influenced by a scalar field $G_{r},$ a qauge field $F^{rs},$ and geodesic motion with respect to the 
Levi-Civita connection corresponding to $g_{ij}.$ The structural appearance of all of these physical aspects is a mathematical
consequence of the choice of non-commutative framework.
\bigbreak

\noindent{\bf Remark.} \noindent It follows from the Jacobi identity that 
$$F_{ij}=g_{ir}g_{js}F^{rs}$$
satisfies the equation 
$$\partial_{i}F_{jk} + \partial_{j}F_{ki} + \partial_{k}F_{ij} = 0,$$
identifying $F_{ij}$ as a non-commutative analog of a gauge field. $G_{i}$
is a non-commutative analog of a scalar field.  The derivation in this section generalizes the Feynman-Dyson derivation of
non-commutative electromagnetism \cite{Dyson} where $g_{ij} = \delta_{ij}.$ In the next section we will say more about the
Feynman-Dyson result. The results of this section sharpen considerably an approach of Tanimura \cite{Tanimura}. In Tanimura's paper,
normal ordering  techniques are used to handle the algebra. In the derivation given above, we have used straight non-commutative
algebra, just as in the original Feynman-Dyson derivation.
\bigbreak

\noindent {\bf Remark.} It is interesting to note that we can rewrite the equation 
$$\ddot{X_{r}} = G_{r} + F_{rs}\dot{X^{s}} + \Gamma_{rst}\dot{X^{s}}\dot{X^{t}}$$ as
$$\ddot{X_{r}} = G_{r} + [\dot{X_{r}}, \dot{X_{s}}]\dot{X^{s}} + \Gamma_{srt}\dot{X^{s}}\dot{X^{t}}.$$
(Just substitute the expression for $F_{rs}$ and recollect the terms.) The reader may enjoy trying her hand at other ways to reorganize
this data. It is important to note that in the first form of the equation, the basic terms $G_{r}$, $F_{rs}$ and  $\Gamma_{rst}$
commute with the coordinates $X_{k}.$ It is this decomposition into parts that commute with the coordinates that guides the structure
of this formula in the non-commutative context.
\bigbreak

\section{An Abstract Version of the Feynman-Dyson Derivation}
In this section we assume that specific time-varying coordinate elements $X_{1},X_{2},X_{3}$ of the algebra $\cal{A}$ are given.
{\it We do not assume any commutation relations about $X_{1},X_{2},X_{3}.$}
We define the field $$H = \dot{X} \times \dot{X}.$$ 
The field $H$ is an analog of the magnetic field in electromagnetic theory and should
not be confused with our earlier notation for the Hamiltonian.
\bigbreak

In this section we no longer avail ourselves of the principle of commutativity that is in back of the original Feynman-Dyson
derivation. (See the last section.) We do not base the derivation to follow on any particular commutation relations about the
variables $X_{i},$ but we do take the definitions of the derivations that we use from that previous context. Surprisingly, the result
is very similar to the one of Feynman and Dyson, as we shall see.
\bigbreak

Here $A \times B$ is the non-commutative vector cross product:
$$(A \times B)_{k} = \Sigma_{i,j = 1}^{3} \epsilon_{ijk}A_{i}B_{j}.$$ (We will drop this summation sign
for vector cross products from now on.) 
Then $$H_{k} = \epsilon_{ijk}\dot{X_{i}}\dot{X_{j}}  = (1/2)\epsilon_{ijk}[\dot{X_{i}},\dot{X_{j}}].$$
We define the field $E$ by the equation $$\ddot{X} = E + \dot{X} \times H.$$
We will see that $E$ and $H$ obey a 
generalization of the Maxwell Equations, and that this generalization describes specific discrete models.
The reader should note that this means that a significant part of the {\it form} of electromagnetism is
the consequence of choosing three coordinates of space, and the definitions of spatial and temporal derivatives with respect to them.
The background process that is being described is otherwise aribitrary, and yet appears to obey physical laws once these
choices are made.
\bigbreak

\noindent {\bf Remarks on the Derivatives.}
\begin{enumerate}
\item Since we do not assume that $[X_{i}, \dot{X_{j}}] = \delta_{ij},$ nor do we assume $[X_{i},X_{j}]=0,$ it will not follow that
$E$ and $H$ commute with the $X_{i}.$ 

\item We continue to define $$\partial_{i}(F) = [F, \dot{X_{i}}],$$ and the reader should note
that, these spatial derivations are no longer flat in the sense of section 4 (nor were they in the original Feynman-Dyson derivation).

\item We define $\partial_{t} = \partial/\partial t$ by the equation
$$\partial_{t}F = \dot{F} - \Sigma_{i}\dot{X_{i}}\partial_{i}(F) =  \dot{F} - \Sigma_{i} \dot{X_{i}}[F, \dot{X_{i}}]$$
for all elements or vectors of elements $F.$ We take this equation as the global definition
of the temporal partial derivative, even for elements that are not commuting with the $X_{i}.$ This notion of temporal partial
derivative
$\partial_{t}$ is a least relation that we can write to describe the temporal relationship of an arbitrary non-commutative vector
$F$ and the non-commutative coordinate vector $X.$

\item In defining $$\partial_{t}F = \dot{F} - \Sigma_{i}\dot{X_{i}}\partial_{i}(F),$$ we
are using the definition itself to obtain a notion of the variation of $F$ with respect to time. The definition itself creates a
distinction between space and time in the non-commutative world.  

\item The reader will have no difficulty verifying the following formula:
$$\partial_{t}(FG) = \partial_{t}(F)G + F\partial_{t}(G) + \Sigma_{i}\partial_{i}(F)\partial_{i}(G).$$
This formula shows that $\partial_{t}$ does not satisfy the Leibniz rule in our non-commutative context.
This is true for the original Feynman-Dyson context, and for our generalization of it. All derivations in this theory that are defined
directly as commutators do satisfy the Leibniz rule. Thus $\partial_{t}$ is an operator in our theory that does not have a
representation as a commutator.

\item We define divergence and curl by the equations
$$\nabla \bullet H = \Sigma_{i=1}^{3} \partial_{i}(H_{i})$$ and 
$$(\nabla \times E)_{k} = \epsilon_{ijk}\partial_{i}(E_{j}).$$
\end{enumerate}
\bigbreak

We now prove a few useful formulas about the vector products.
First we have the basic identity about the epsilon.
\bigbreak

\noindent {\bf Lemma 7.1} Let $\epsilon_{ijk}$ be the epsilon tensor taking values $0$, $1$ and $-1$ as follows: When $ijk$ is a
permuation of $123$, then $\epsilon_{ijk}$ is equal to the sign of the permutation. When $ijk$ contains a repetition from 
$\{1,2,3 \},$ then the value of epsilon is zero. 
Then $\epsilon$ satisfies the following identity in terms of the Kronecker delta. 
$$\Sigma_{i} \,\epsilon_{abi}\epsilon_{cdi} =  -\delta_{ad}\delta_{bc} + \delta_{ac}\delta_{bd}.$$
\bigbreak

\noindent The proof of this identity is left to the reader. The identity itself will be referred to as the {\em epsilon identity}.
The epsilon identity is a key structure in the work of this section, and indeed in all formulas involving the vector cross product.
\bigbreak

\noindent {\bf Lemma 7.2} Let $A, B, C$ be vectors of elements of the algebra $\cal{A}.$
Then $$(A \times B) \bullet C = A \times (B \bullet C).$$
\bigbreak

\noindent{\bf Proof.} Note that 
$$(A \times B) \bullet C = \Sigma_{ijk}\, \epsilon_{ijk} A_{i} B_{j} C_{k}$$
$$= \Sigma_{ijk}\, A_{i} \epsilon_{ijk} B_{j} C_{k} $$
$$= \Sigma_{ijk}\, A_{i} \epsilon_{jki} B_{j} C_{k} $$
$$= A \bullet (B \times C).$$
This completes the proof of the Lemma. $\hfill \Box$
\bigbreak

\noindent {\bf Lemma 7.3} Let $A$ be any vector of three elements of the algebra $\cal{A}.$ Then 
$$\nabla \times A = - A \times \dot{X}  - \dot{X} \times A.$$
\bigbreak

\noindent {\bf Proof.} We shall use the summation convention for repeated indices in this calculation.
$$(\nabla \times A)_{k} = \epsilon_{ijk} \partial_{i} A_{j} = \epsilon_{ijk} [A_{j}, \dot{X_{i}}]$$
$$ = \epsilon_{ijk} A_{j} \dot{X_{i}} - \epsilon_{ijk} \dot{X_{i}} A_{j}$$
$$ = - \epsilon_{jik} A_{j} \dot{X_{i}} - \epsilon_{ijk} \dot{X_{i}} A_{j}$$
$$ = - (A \times \dot{X})_{k} - (\dot{X} \times A)_{k}.$$
This completes the proof of the Lemma. $\hfill \Box$
\bigbreak

\noindent {\bf Lemma 7.4} For $A$ and $B$ any elements in the algebra $\cal{A},$
$$\nabla \times (A \times B) = -(\nabla \bullet A)B - A \bullet(\nabla B) + (\nabla A)\bullet B + A(\nabla \bullet B)$$
where $$(A \bullet(\nabla B))_{i} = \Sigma_{k} A_{k}\partial_{k}B_{i}$$ and
$$(((\nabla A)\bullet B)_{i} = \Sigma_{k} (\partial_{k}A_{i})B_{k})_{c}.$$
\bigbreak

\noindent {\bf Proof.} $$(\nabla \times (A \times B))_{c}  = \epsilon_{abi}\epsilon_{cdi} \partial_{d}(A_{a}B_{b})$$
$$= (-\delta_{ad}\delta_{bc} + \delta_{ac}\delta_{bd}) \partial_{d}(A_{a}B_{b})$$
$$= -\delta_{ad}\delta_{bc}\partial_{d}(A_{a}B_{b}) + \delta_{ac}\delta_{bd} \partial_{d}(A_{a}B_{b})$$
$$= -\partial_{a}(A_{a}B_{c}) + \partial_{b}(A_{c}B_{b})$$
$$= -\partial_{a}(A_{a})B_{c} -A_{a}\partial_{a}(B_{c}) + \partial_{b}(A_{c})B_{b} + A_{c} \partial_{b}(B_{b})$$
$$= [-(\nabla \bullet A)B - A \bullet(\nabla B) + (\nabla A)\bullet B + A(\nabla \bullet B)]_{c}.$$
This completes the proof of the Lemma. $\hfill \Box$
\bigbreak

\noindent {\bf Remark.} This Lemma, and the observation that the formula in the Lemma works in the non-commutative context is
due to the author and Keith Bowden in conversations around $1999.$ See \cite{KB}. We now give the generalization of the Feynman-Dyson
result in  this formalism.
\bigbreak

\noindent {\bf Theorem 7.5} With the above definitions of the operators, and taking
$$\nabla^{2} = \partial_{1}^{2} + \partial_{2}^{2} + \partial_{3}^{2}, \,\,\, H = \dot{X} \times \dot{X} \,\,\, \mbox{and} \,\,\, E =
\partial_{t}\dot{X} \,\,\, \mbox{we have}$$

\begin{enumerate}
\item $\ddot{X} = E + \dot{X} \times H$
\item $\nabla \bullet H = 0$
\item $\partial_{t}H + \nabla \times E = H \times H$
\item $\partial_{t}E - \nabla \times H = (\partial_{t}^{2} - \nabla^{2})\dot{X}$
\end{enumerate}
\bigbreak

\noindent {\bf Remark.} Note that this Theorem is a non-trivial generalization of the Feynman-Dyson derivation of electromagnetic 
equations. In the Feynman-Dyson case, one assumes that the commutation relations
$$[X_{i}, X_{j}] = 0$$ and 
$$[X_{i}, \dot{X_{j}}] = \delta_{ij}$$ are given, {\em and} that the principle of commutativity is in place, so that 
if $A$ and $B$ commute with the $X_{i}$ then $A$ and $B$ commute with each other. One then can interpret $\partial_{i}$ as a 
standard derivative with $\partial_{i}(X_{j}) = \delta_{ij}.$ Furthermore, one can verify that $E_{j}$ and $H_{j}$ both commute with
the $X_{i}.$ From this it follows that $\partial_{t}(E)$ and $\partial_{t}(H)$ have standard intepretations and that $H \times H = 0.$
The above formulation of the Theorem adds the description of $E$ as $\partial_{t}(\dot{X}),$ a non-standard use of 
$\partial_{t}$ in the original context of Feyman-Dyson, where $\partial_{t}$ would only be defined for those $A$ that commute with 
$X_{i}.$ In the same vein, the last formula $\partial_{t}E - \nabla \times H = (\partial_{t}^{2} - \nabla^{2})\dot{X}$ gives a way
to express the remaining Maxwell Equation in the Feynman-Dyson context.
\bigbreak

\noindent {\bf Proof of Theorem.} We begin  by calculating
$$W = \dot{X} \times H = \dot{X} \times (\dot{X} \times \dot{X}).$$ Hence
$$W_{i} = (\dot{X} \times (\dot{X} \times \dot{X}))_{i} = - \dot{X_{k}}(\dot{X_{k}}\dot{X_{i}}) +
(\dot{X_{k}}\dot{X_{i}})\dot{X_{k}}.$$ This follows from Lemma 7.1. Hence
$$W_{i} = [\dot{X_{k}}\dot{X_{i}},\dot{X_{k}}] = \Sigma_{k} \partial_{k}(\dot{X_{k}}\dot{X_{i}})$$
$$= \Sigma_{k} \partial_{k}(\dot{X_{k}}) \dot{X_{i}} + \dot{X_{k}} \partial_{k}(\dot{X_{i}}).$$
But $$\partial_{k}(\dot{X_{k}}) = [\dot{X_{k}}, \dot{X_{k}}] = 0,$$ hence
$$W_{i} = \Sigma_{k} \dot{X_{k}} \partial_{k}(\dot{X_{i}}) = \ddot{X_{i}} - \partial_{t}\dot{X_{i}} = \ddot{X_{i}} - E_{i}.$$
Thus $$\dot{X} \times H = \ddot{X} - E.$$ This completes the proof of the first part.
\smallbreak

$$\nabla \bullet H = \Sigma_{i} [H_{i}, \dot{X_{i}}] = H \bullet \dot{X} - \dot{X} \bullet H$$
$$= (\dot{X} \times \dot{X})  \bullet \dot{X} - \dot{X} \bullet (\dot{X} \times \dot{X}) = 0,$$
since it is easy to verify that $(A \times B) \bullet C = A \bullet (B \times C)$ for the non-commutative vector cross product.
\bigbreak

\noindent Since $$\partial_{t}H = \dot{H} - \dot{X} \bullet(\nabla H),$$ we have
$$\partial_{t}H + \nabla \times E = \dot{H} - \dot{X} \bullet (\nabla H) + \nabla \times E$$
$$= \dot{(\dot{X} \times \dot{X})} - \dot{X} \bullet (\nabla H) + \nabla \times E$$
$$= \ddot{X} \times \dot{X} + \dot{X} \times \ddot{X} - \dot{X} \bullet (\nabla H) + \nabla \times E$$
$$= (E + \dot{X} \times H) \times \dot{X} + \dot{X} \times (E + \dot{X} \times H) - \dot{X} \bullet (\nabla H) + \nabla \times E$$
$$= (E \times \dot{X} + \dot{X} \times E) + (\dot{X} \times H) \times \dot{X} + \dot{X} \times (\dot{X} \times H) - \dot{X} \bullet
(\nabla H) + \nabla \times E$$
$$= - \nabla \times E + (\dot{X} \times H) \times \dot{X} + \dot{X} \times (\dot{X} \times H) - \dot{X} \bullet
(\nabla H) + \nabla \times E$$
$$= (\dot{X} \times H) \times \dot{X} + \dot{X} \times (\dot{X} \times H) - \dot{X} \bullet
(\nabla H) $$
$$= - \nabla \times (\dot{X} \times H) - \dot{X} \bullet (\nabla H). $$
Now, using the formula for $\nabla \times (A \times B),$ we obtain
$$\partial_{t}H + \nabla \times E 
= (\nabla \bullet \dot{X})H + \dot{X} \bullet(\nabla H) -  (\nabla \dot{X})\bullet H - \dot{X}(\nabla \bullet H)
 - \dot{X} \bullet (\nabla H)$$
$$ = (\nabla \bullet \dot{X})H  -  (\nabla \dot{X})\bullet H - \dot{X}(\nabla \bullet H).$$
Note that $\nabla \bullet \dot{X} = \Sigma_{i}[\dot{X_{i}},\dot{X_{i}}] = 0$ and that $\nabla \bullet H = 0.$ Thus
$$\partial_{t}H + \nabla \times E = -  (\nabla \dot{X})\bullet H.$$  
Now note that $$(H \times H)_{k} = ((\dot{X} \times \dot{X}) \times H)_{k} = \epsilon_{ijk}(\dot{X} \times \dot{X})_{i}H_{j}
= \epsilon_{ijk}\epsilon_{rsi}\dot{X_{r}}\dot{X_{s}}H_{j}$$
$$= (-\delta_{js}\delta_{kr} + \delta_{jr}\delta_{ks})\dot{X_{r}}\dot{X_{s}}H_{j}$$
$$= -\dot{X_{k}}\dot{X_{j}}H_{j} + \dot{X_{j}}\dot{X_{k}}H_{j}$$
$$= -\Sigma_{j} [\dot{X_{k}}, \dot{X_{j}}]H_{j} = (-(\nabla \dot{X})\bullet H)_{k}.$$
Therefore $$\partial_{t}H + \nabla \times E = H \times H.$$ 
\bigbreak

\noindent Now consider 
$$(\nabla \times H)_{i} = -\partial_{k}(\dot{X_{k}}\dot{X_{i}}) + \partial_{k}(\dot{X_{i}}\dot{X_{k}})$$
$$ = \partial_{k}[\dot{X_{i}}, \dot{X_{k}}] = \Sigma_{k} [[\dot{X_{i}}, \dot{X_{k}}], \dot{X_{k}}]$$
$$= \Sigma_{k}\partial_{k}^{2}\dot{X_{i}}.$$ Hence
$$\nabla \times H = \nabla^{2}\dot{X}.$$
The last part of the Theorem follows immediately from this formula.
This completes the proof.   $\hfill \Box$
\bigbreak 

\noindent {\bf Remark.} Note the role played by the epsilon tensor $\epsilon_{ijk}$ throughout the construction of 
generalized electromagnetism in this section. The epsilon tensor is the structure constant for the Lie algebra of the rotation
group $SO(3).$ If we replace the epsilon tensor by a structure constant
$f_{ijk}$ for a Lie algebra ${\cal G}$of dimension $d$ such that the tensor is invariant under cyclic permutation ($f_{ijk} =
f_{kij}$), then most of the  work in this section will go over to that context. We would then have $d$ operator/variables $X_1,
\cdots X_d$ and a generalized  cross product defined on vectors of length $d$ by the equation
$$(A \times B)_{k} = f_{ijk}A_{i}B_{j}.$$
The Jacobi identity for the Lie algebra ${\cal G}$ implies that this cross product will satisfy
$$A \times (B \times C) = (A \times B) \times C + [B \times (A ] \times C)$$
where $$([B \times (A ] \times C)_{r} = f_{klr}f_{ijk}A_{i}B_{k}C_{j}.$$ This extension of the Jacobi identity 
holds as well for the case of non-commutative cross product defined by the epsilon tensor.
The reader will enjoy looking back over this section and seeing that we can still carry Theorem 7.5
up to the following conclusion with $E$ defined by the second equation below. We can no longer take $E = \partial_{t}\dot{X},$ as this
depends upon the specific properties of the epsilon. 
\begin{enumerate}
\item Assume $H = \dot{X} \times \dot{X}$.
\item Assume $\ddot{X} = E + \dot{X} \times H$.
\item Then $\nabla \bullet H = 0$.
\item Then $\partial_{t}H + \nabla \times E = - \nabla \times (\dot{X} \times H) - \dot{X} \bullet (\nabla H)$.
\end{enumerate}
It is therefore of interest to explore the structure of generalized non-commutative electromagnetism over other Lie algebras
(in the above sense). This will be the subject of another paper.
\bigbreak

 \subsection{The Original Feynman - Dyson Derivation and its Gauge Theoretic Context}
The original Feynman-Dyson derivation \cite{Dyson, KN:QEM, Hughes, Mont} assumes that we have three variables
$\{X_1,X_2,X_3\}$ and the commutation relations
$$[X_i,X_j]=0$$
$$[X_i, \dot{X_j}] = \delta_{ij}.$$
It is also assumed that if $A$ and $B$ commute with the $X_i$, then $A$ and $B$ commute with each other. That is, $A$ and $B$
are then ``functions of the $X_i$". We have called this the principle of commutativity.
\bigbreak

\noindent With these assumptions one proves that with 
$$H = \dot{X} \times \dot{X},$$ (non-commutative  vector cross product) and $E$ defined by $$\ddot{X} = E + \dot{X} \times H,$$ then
$E$ and $H$ satisfy the Maxwell equations in the sense that 
\begin{enumerate}
\item $E$ and $H$ commute with the $X_{i}.$
\item $\nabla \bullet H = 0.$
\item $\partial_{t}H + \nabla \times E = 0.$
\end{enumerate}
where these differential operators have been described in detail (in the non-commutative framework) in this section.
A key to the original demonstration is the principle of commutativity, providing a compass for comparing the results with the context
of classical calculus. In this section we have seen that an abstraction of the Feynman-Dyson argumemt provides a serious 
generalization that encompasses a number of discrete models (to be discussed below). In this sub-section, we compare the Feyman-Dyson
framework with our  already-constructed formality of non-commutative gauge theory.
\bigbreak 

We use the dynamics $$dX_{i}/dt = \dot{X_{i}} = P_{i} - A_{i},$$ as before.
We restrict to the case where $[X_{i}, A_{j}]=0$ so that 
$$g_{ij} = [X_{i}, \dot{X_{j}}] = [X_{i}, P_{j} - A_{j}] = \delta_{ij} - [X_{i}, A_{j}] = \delta_{ij}.$$
This is the domain to which the original Feynman-Dyson derivation applies.  

\noindent We then have

$$[X_i,X_j]=0$$
$$[X_i, \dot{X_j}] = \delta_{ij}$$

\noindent and 

$$R_{ij} = [\dot{X_i}, \dot{X_j}] = \partial_{i}A_j - \partial_{j}A_i + [A_i, A_j].$$

\noindent Note that even under these restrictions we are still looking at the possibility of a
non-abelian gauge field. The pure electromagnetic case is when the commutator of $A_i$ and 
$A_j$ vanishes.  In the Feynman-Dyson context, this commutator does vanish, since it is given that 
$[X_{i}, A_{j}]=0$ for all $i$ and $j,$ and the principle of commutativity  applies.
\bigbreak

With this interpretation, $E$ is defined by 
the Lorentz force law $$\ddot{X} = E + \dot{X} \times H$$ where $H$ represents the 
magnetic field. To see how this works, suppose that
$\ddot{X_i} = E_i +F_{ij}\dot{X_j}$ and  suppose that $E_i$ and $F_{ij}$ commute with $X_k.$ Then we can compute
$$[X_i,\ddot{X_j}] = [X_i, E_j +F_{jk}\dot{X_k}]$$
$$= F_{jk}[X_i, \dot{X_k}] = F_{jk}\delta_{ik} = F_{ji}.$$

\noindent This implies that $$F_{ij} = [\dot{X_i}, \dot{X_j}] = R_{ij} = \partial_{i}A_j - \partial_{j}A_i$$

\noindent since $[X_i,\ddot{X_j}] + [\dot{X_i}, \dot{X_j}] = D[X_i, \dot{X_j}] = 0.$
It is then easy to verify that the Lorentz force equation is satisfied with 
$H_k = \epsilon_{ijk}R_{ij}$ and that in this case of $[A_i,A_j]=0$ leads directly to 
standard electromagnetic theory when the bracket is a Poisson bracket. When this bracket is not zero but the
potentials $A_i$ are functions only of the $X_j$ we can look at a generalization of gauge theory where the 
non-commutativity comes from internal Lie algebra parameters. This shows how a shift of the original Feynman-Dyson 
derivation supports generalizations of classical electromagnetism.
\bigbreak

\subsection{Discrete Thoughts}
In the hypotheses of the above Theorem, we are free to take any non-commutative world, and
the Theorem will
satisfied in that world. For example, we can take each $X_{i}$ to be an arbitary time series of real or complex numbers, or
bitstrings of zeroes and ones. The global time derivative is defined by $$\dot{F} = J(F' - F) = [F, J],$$ where $FJ =
JF'.$ This is the non-commutative discrete context discussed in sections 2 and 3. We will write
$$\dot{F} = J\Delta(F)$$ where $\Delta(F)$ denotes the classical discrete derivative
$$\Delta(F) = F' -F.$$ 
With this interpretation
$X$ is a vector with three real or complex coordinates at each time, and  
$$H = \dot{X} \times \dot{X} = J^{2}\Delta(X') \times \Delta(X)$$ while
$$E = \ddot{X} - \dot{X} \times (\dot{X} \times \dot{X}) = J^{2}\Delta^{2}(X) - J^{3} \Delta(X'') \times ( \Delta(X') \times
\Delta(X)).$$ Note how the non-commutative vector cross products are composed through time shifts in this context of temporal
sequences of scalars. The advantage of the generalization now becomes apparent. We can create very simple models of generalized
electromagnetism with only the simplest of discrete materials. In the case of the model in terms of triples of time series, the 
generalized electromagnetic theory is a theory of measurements of the time series whose key quantities are
$$\Delta(X') \times \Delta(X)$$ and 
$$\Delta(X'') \times (\Delta(X') \times \Delta(X)).$$
\bigbreak

It is worth noting the forms of the basic derivations in this model. We have, assuming that $F$ is a commuting scalar (or vector of
scalars) and taking $\Delta_{i} = X_{i}' - X_{i},$
$$\partial_{i}(F) = [F, \dot{X_{i}}] =[F, J\Delta_{i}] = FJ\Delta_{i} - J\Delta_{i}F = J(F'\Delta_{i} - \Delta_{i}F)
= \dot{F}\Delta_{i}$$  and for the temporal derivative we have
$$\partial_{t}F = J[1 - J \Delta' \bullet \Delta]\Delta(F)$$ where
$\Delta = (\Delta_{1}, \Delta_{2}, \Delta_{3}).$
\bigbreak

\subsubsection{Discrete Classical Electromagnetism}
It is of interest to compare these results with a direct discretization of classical electromagnetism.
Suppose that $X,X',X'',X''', \cdots$ is a time series of vectors in $R^{3}$ (where $R$ denotes the real numbers).
Let ${\cal D}X = X' - X$ be the usual discrete derivative (with time step equal to one for convenience).
Let $A \bullet B$ denote the usual inner product of vectors in three dimesions.

{\it Assume that there are fields $E$ and $H$ such that $$ {\cal D}^{2}X = E + {\cal D}X \times H$$ 
(the Lorentz force law). Assume that $E$ and $H$ are perpendicular to the velocity vector ${\cal D}X$, and 
that $E$ is perpendicular to $H.$ } 
\bigbreak

\noindent Then we have
$${\cal D}X' \times {\cal D}X = ({\cal D}X' - {\cal D}X) \times {\cal D}X = ({\cal D}^{2}X) \times ({\cal D}X)$$
$$= E \times {\cal D}X + ({\cal D}X \times H) \times {\cal D}X$$
$$= E \times {\cal D}X - {\cal D}X (H \bullet {\cal D}X) + ({\cal D}X \bullet {\cal D}X)H.$$ 
Since $E$ is perpendicular to ${\cal D}X$ we know there is a $\lambda$ such that $E \times {\cal D}X = \lambda H$
and we have $H \bullet {\cal D}X = 0$ since $H$ is perpendicular to ${\cal D}X.$ Therefore
$${\cal D}X' \times {\cal D}X = \lambda H + ||{\cal D}X||^{2} H$$  so that
$$H = {\cal D}X' \times {\cal D}X /(\lambda + ||{\cal D}X||^{2}).$$ 
\bigbreak

\noindent Now using $\Delta = \Delta(X),$ and get
$$E = (\Delta' - \Delta)  - [\Delta \times (\Delta' \times \Delta)]/(\lambda + \Delta \bullet \Delta),$$ and 
$$H = (\Delta' \times \Delta)/(\lambda + \Delta \bullet \Delta).$$
 
\noindent The formula for $H$ is in exactly the same pattern as the formula for $H$ in the discrete model for
generalized electromagnetism as described in this subsection.  Up to the time-shifting
algebra and a proportionality constant, the expressions are the same. The expression for $E$ is similar, but involves different
time-shift structure.  Clearly more work is needed in comparing classical discrete electromagnetism
with the results of a discrete analysis of this generalized Feynman-Dyson derivation.
\bigbreak

\subsection{More Discrete Thoughts}
In the Feynman-Dyson derivation of electromagnetic
formalism from commutation relations \cite{KN:QEM} one uses the relations 
$$[X_i, X_j] = 0$$
$$[X_i, \dot{X_j}] = k\delta_{ij}$$
where $k$ is a scalar.
In this subsection we shall use
$$[X_i, X_j] = 0$$
$$[X_i, \dot{X_j}] = Jk\delta_{ij}$$ as we did in analyzing the one-dimensional case.
We shall take $$\dot{F} = J(F' -F) = [F,J]$$ with $$JF=F'J,$$ taking the time-step equal to one for convenience.
This allows us to have scalar evolution of the time series, but changes the issues in the original Feynman-Dyson
derivation due to presence of the non-commutative operator $J$ in the second equation. These issues  are handled by the more
general formalism that we discussed in this section. We aim to see to what extent one can make
simple models for this version of the  Feynman-Dyson relations. Models of this sort will be another level of approximation to 
discrete  electromagnetism. 
\bigbreak 

Writing out the commutation relation $[X,\dot{X}]=Jk$, and not making
any assumption that $X'$ commutes with $X$, we find
$$J^{-1}[X,\dot{X}] = X'(X'-X) -(X'-X)X$$
$$= X'(X'-X) -X(X'-X) +X(X'-X) -(X'-X)X$$
$$= (X'-X)^2 + (XX'-X'X) = (X'-X)^2 + [X,X'].$$

\noindent Thus the commutation relation $[X,\dot{X}]=Jk$ becomes the equation
$$(X'-X)^2 + [X,X'] = k.$$

\noindent By a similar calculation, the equation $[X,\dot{Y}]=0$ becomes the equation
$$(X'-X)(Y'-Y) + [X,Y']=0.$$

These equations are impossible to satisfy simultaneously for $k \ne 0$ 
if we assume that $X$ and $X'$ commute and that 
$X$ and $Y'$ commute and that $[Y,\dot{Y}]=Jk$. For then we would need to solve:
$$(X'-X)^2 = k.$$
$$(Y'-Y)^2 = k.$$
$$(X'-X)(Y'-Y) =0.$$
\noindent with the first two equations implying that $(X-X')$ and $(Y-Y')$ are each
non-zero, and the third implying that their product is equal to zero.  
In other words, the equations below cannot be satisfied if the time series are composed of
commuting scalars. 
$$[X,\dot{X}]=Jk$$
$$[Y,\dot{Y}]=Jk$$
$$[X,Y]=0$$ 
{\em In order to make such models we shall have to introduce non-commutativity into
the time series themselves.}  
\bigbreak

\noindent Here is an example of such a model.
\smallbreak
\noindent Return to the equations
$$(X'-X)^2 + [X,X'] = k.$$
$$(X'-X)(Y'-Y) + [X,Y']=0$$ \noindent expressing the behaviour for two distinct variables $X$ 
and $Y.$ If $[X,X'] = 0$, then we have $(X'-X)^2 = k$ so that 
$$X' = X \pm \sqrt{k}.$$  

\noindent In order for the second equation to be satisfied, we need that 
$$[X,Y'] = \pm k$$ \noindent where the ambiguity of sign is linked with the varying signs in the 
temporal behaviour of $X$ and $Y.$ We will make the sign more precise in a moment, but {\em the radical 
part of this suggestion is that for two distinct spatial variables $X$ and $Y$, there will be a 
commutation relation between one and a time shift of the other.}
\bigbreak

\noindent If the space variables are labeled $X_i$, then we can write 
$$X_{i}^{t+1} = X_{i}^{t} + \epsilon_{i}^{t} k$$ \noindent where $\epsilon_{i}^{n}$
is plus one or minus one. Thus each space variable performs a walk with the fixed step-length
$k.$  We shall write informally $$X_{i}' = X_{i} + \epsilon_{i} k$$ \noindent where it is 
understood that the epsilon without the superscript connotes the sign change that occurs in this
juncture of the process. We then demand the commutation relations
$$[X_{i}', X_{j}] = [X_{j}', X_{i}] = \epsilon_{i} \epsilon_{j} k.$$ \noindent Each $X_{i}$ is
a scalar in its own domain, but does not commute with the time shifts of the other directions.
We then can have the full set of commutation relations:
$$[X_{i}', X_{j}] = [X_{j}', X_{i}] = \epsilon_{i} \epsilon_{j} k.$$
$$[X_i, X_j] = 0$$
$$[X_i, \dot{X_j}] = Jk\delta_{ij}.$$   In this system,the elements of a given time series
$X_{i}, X_{i}',X_{i}'', \cdots$ commute with one another. The basic field element in the generalized Feynman-Dyson
set up is the magnetic field $H$ defined by the (non-commutative) vector cross product
$$H =(1/k)  \dot{X} \times \dot{X}.$$ Here we have
$$\dot{X_{i}} = J(X_{i}' - X_{i}) = J \epsilon_{i} \sqrt{k}.$$ Thus $$H = J^{2} ~ \epsilon' \times \epsilon$$
\noindent where $\epsilon =(\epsilon_{1}, \epsilon_{2}, \epsilon_{3})$  and 
$\epsilon'$ denotes this vector of signs at the next time step. In this way we see that we can think of each
spatial coordinate as providing a long temporal bit string and the three coordinates together give the field in
terms of the vector cross product of their temporal cross sections at neighboring instants. It is interesting to
compare this model with the color algebra in \cite{Wene}. 
\bigbreak

\section{The Jacobi Identity and Poisson Brackets}
It is worth thinking through the message of the non-commutative world in respect to the existence of the Poisson brackets
and their connection with continous differentiation and the commutative world of topology and differential geometry from which
the classical and the quantum models of physics are derived. In the classical world there are specific point locations, and the notion
of a trajectory is given in terms of a continuous sequence of such locations. But there is no inherent operational structure intrinsic 
to the space. There is great freedom in the world of commutative and continuous calculus, a freedom that allows the construction
of many models of temporal evolution. Yet we have seen that non-commutative worlds have built in laws, and built in patterns of 
evolution. These patterns of evolution do not lead directly to trajectories but rather to patterns of concatenations of 
operators. At first sight it would seem that there could be no real connection between these worlds. The Poisson bracket
and the reformulation of mechanics in Hamiltonian form shows that this is not so. There is a special non-commutativity inherent
in the continuous calculus, via the Poisson Bracket.
\bigbreak

It is easy to see the truth of the Jacobi identity for commutators.
It is just a little harder to see the Jacobi identity of Poisson brackets.
It is the purpose of this section to recall these verifications and to discuss the nature of the 
identity.
\bigbreak

First let $[A,B] = AB-BA.$ Then 
$$[[A,B],C] = (AB-BA)C - C(AB-BA) = ABC -BAC -CAB + CBA.$$
$$[[A,B],C] = ABC - BAC - CAB + CBA,$$
$$[[C,A],B] = CAB - ACB - BCA + BAC,$$
$$[[B,C],A] = BCA - CBA - ABC + ACB.$$ So
$$[[A,B],C] + [[C,A],B]+ [[B,C],A] = 0.$$ This is the {\em Jacobi identity}.
\bigbreak

More generally, a Lie algebra is an algebra 
$\cal A$ with a (non-associative) product
$ab$, not necessarily a commutator, that satisfies
\begin{enumerate}
\item Jacobi identity $(ab)c + (bc)a + (ca)b = 0$ and 
\item ba = -ab.
\end{enumerate}
It follows that if we define $\rho_{a}: \cal A \longrightarrow \cal A$ by the
equation $\rho_{a}(x) = ax$ for each $a$ in $\cal A,$ then 
$$\rho_{ab} = [\rho_{a}, \rho_{b}],$$ so that products go to commutators
naturally in  the left-regular representation of the algebra upon itself.  
\bigbreak

\noindent Here is another point of view. We have the following equivalent form of the Jacobi identity
(when $ab=-ba$ for all $a$ and $b$):
$$a(xy) = (ax)y + x(ay)$$
for all $a$, $x$ and $y$ in the algebra. This identity says that each element $a$ in the algebra acts, by left multiplication, as
a derivation on the algebra.  In this way, we see that Lie algebras are the natural candidates as contexts for non-commutative 
worlds that contain an image of the calculus. 
\bigbreak

\subsection{Proving the Jacobi Identity for Poisson Brackets}
There are examples of Lie algebras where the non-associative product is not a commutator, 
the most prominent being the {\em Poisson bracket}. Here we start with a {\em commutative} algebra $\cal
CA$ with two (or more) derivations on $\cal CA.$  Let there be operators $\UL{a}$ and $\UR{b}$ acting on 
$\cal CA$ ($ab$ is the commutative multiplication) such that these operators satisfy
the Leibniz rule and commute with one another:
$$\UL{ab} = \UL{a}\,b + a\, \UL{b}$$ and 
$$\UR{ab} = \UR{a}\,b + a\, \UR{b},$$ and 
$$\UL{\UR{a}} = \UR{\UL{a}}$$ for all elements of $\cal CA$. 
Then we define the Poisson Bracket on $\cal CA$ by the formula
$$\{a,b\}_{\cal CA} = \UL{a} \UR{b} - \UL{b} \UR{a}.$$
We wish to see that this product satisfies the Jacobi identity. In order to do this we first prove a 
lemma about the Jacobi identity for commutators in a non-associative algebra. We then apply that lemma
to the specific non-associative product $$a*b = \UL{a}\UR{b}.$$
\bigbreak 

Suppose that $*$ denotes a non-commutative and non-associative binary operation.
We want to determine when the commutator $[A,B] = A*B - B*A$ satisfies the Jacobi identity.
We first prove a lemma about the Jacobi identity for commutators in 
a non-associative algebra. Let $\cal NA$ be a non-associative linear algebra with multiplication denoted
by
$*$ as above. Let 
$$J(a,b,c) = [[a,b],c] + [[c,a],b] + [[b,c],a],$$ and call this the {\em Jacobi sum} of $a,b$ and $c.$
We say that the Jacobi identity is satisfied for all elements $a,b,c \in \cal NA$ if $J(a,b,c) = 0$
for all $a,b,c \in \cal NA.$ We define the {\em associator} of elements $a,b,c$ by the formula
$$<a,b,c> = (a*b)*c - a*(b*c).$$ Let $\sigma$ be an element of the permutaion group $S_3$ on three
letters, acting on the set $\{a,b,c\}.$  Let $a^{\sigma},b^{\sigma},c^{\sigma}$ be the images of 
$a,b,c$ under this permutation.  Let $sgn(\sigma)$ denote the sign of the permutation.
\bigbreak

\noindent {\bf Lemma 8.1} Let $\cal NA$ be a non-associative algebra as above, then the the Jacobi sum 
$J(a,b,c) = [[a,b],c] + [[c,a],b] + [[b,c],a],$ for any elements $a,b,c \in A$ is given by the formula
$$J(a,b,c) = \Sigma_{\sigma \in S_3} sgn(\sigma)<a^{\sigma},b^{\sigma},c^{\sigma}>.$$
Thus the Jacobi identity is satisfied in $\cal NA$ iff the following identity is true for all $a,b,c \in
\cal NA.$
$$\Sigma_{\sigma \in S_3} sgn(\sigma)<a^{\sigma},b^{\sigma},c^{\sigma}>=0.$$
\bigbreak 

\noindent {\bf Proof.}  For the duration of this proof we shall write $ab$ for $a*b.$
Then
$$[[a,b],c] = (ab-ba)c - c(ab-ba) = (ab)c -(ba)c -c(ab) + c(ba),$$
$$[[c,a],b] = (ca-ac)b -b(ca-ac)  = (ca)b -(ac)b -b(ca) + b(ac),$$
$$[[b,c],a] = (bc-cb)a -a(bc-cb) = (bc)a - (cb)a -a(bc) + a(cb).$$
Hence
$$[[a,b],c]+[[c,a],b]+[[b,c],a]$$ 
$$= (ab)c -(ba)c + (ca)b -(ac)b + (bc)a - (cb)a$$
$$-c(ab) + c(ba) -b(ca) + b(ac) -a(bc) + a(cb)$$                        

$$= ((ab)c-a(bc)) - ((ba)c - b(ac)) + ((ca)b - c(ab))$$
$$- ((ac)b - a(cb))+((bc)a - b(ca)) -((cb)a - c(ba))$$                                
                                  
$$= <a,b,c> - <b,a,c> + <c,a,b>$$ 
$$- <a,c,b> + <b,c,a> - <c,b,a>.$$
This completes the proof.    
\bigbreak

\noindent {\bf Remark.} We discovered this lemma in the course of the research for this paper. Gregory Wene points out to us 
that a version of the lemma can be found in \cite{Myung}. We now apply this result to prove that Poisson Brackets satisfy the Jacobi
identity.
\bigbreak

\noindent {\bf Theorem 8.2} Let there be operators $\UL{a}$ and $\UR{b}$ acting on a
commutative algebra $\cal CA$ ($ab$ is the commutative multiplication) such that these operators satisfy
the Leibniz rule and commute with one another:
$$\UL{ab} = \UL{a}\,b + a\, \UL{b}$$ and 
$$\UR{ab} = \UR{a}\,b + a\, \UR{b},$$ and 
$$\UL{\UR{a}} = \UR{\UL{a}}$$ for all elements of $\cal CA$. Define a non-associative algebra $\cal NA$
with product
$$a*b = \UL{a}\,\UR{b}.$$ Then the commutator in this algebra $[a,b]_{A} = a*b - b*a$ will satisfy the
Jacobi identity.  Note that this commutator is the Poisson bracket with respect to the above derivations
in the original commutative algebra:
$$\{a,b\}_{\cal CA} = \UL{a}\UR{b} - \UL{b}\UR{a} = a*b-b*a = [a,b]_{\cal NA}.$$
This result implies that Poisson brackets satisfy the Jacobi identity.
\bigbreak

\noindent {\bf Proof.}  Consider the associator in the non-associative algebra defined in the statement
of the Theorem:
$$<a,b,c> = (a*b)*c - a*(b*c) = \UL{\UL{a}\UR{b}}\UR{c} - \UL{a}\UR{\UL{b}\UR{c}}$$
$$= \UL{\UL{a}} \, \UR{b} \, \UR{c} + \UL{a} \, \UL{\UR{b}} \, \UR{c} - \UL{a} \, \UR{\UL{b}} \, \UR{c} +
\UL{a} \, \UL{b} \, \UR{\UR{c}}$$
$$= \UL{\UL{a}} \, \UR{b} \, \UR{c} + \UL{a} \, \UL{b} \, \UR{\UR{c}}$$
Note that an expression of the form
$$\UL{\UL{a}} \, \UR{b} \, \UR{c}$$
will return zero when averaged in the summation 
$$\Sigma_{\sigma \in S_3} sgn(\sigma)<a^{\sigma},b^{\sigma},c^{\sigma}>$$
since $ \UL{\UL{a}} \, \UR{b} \, \UR{c} =\UL{\UL{a}} \, \UR{c} \, \UR{b}$ (the underlying algebra is
commutative) and these terms will appear with opposite signs in the summation. 
Therefore we find that $Jac(a,b,c)=0$ for all $a,b,c$ in $R.$ This completes the proof.
\bigbreak

\section {Diagrammatics and the Jacobi Identity}
We have seen that a commutative world equipped with distinct derivations that commute with one another is sufficient to
produce a non-commutative world (via the Poisson brackets) that is strong enough to support our story of physical patterns. Many
combinatorial patterns mimic the Jacobi identity, and hence provide fuel for further study. In order to
illustrate these connections, we give in this section a diagrammatic version of the Jacobi identity and an interpretation in terms of
graph coloring. We will initially work in an Lie  algebra
$\cal G$ whose product $ab$ satisfies $ba = -ab$ and the Jacobi identity
$a(bc) = (ab)c + a(bc).$ In Figure 3 we show a diagrammatic interpretation of multiplication, consisting in a trivalent vertex
labeled with $a$, $b$, and $ab.$ As one moves around the vertex in the plane, clockwise, one encounters first $a$, then $b$, and 
then $ab.$
\bigbreak

{\tt    \setlength{\unitlength}{0.92pt}
\begin{picture}(213,124)
\thinlines    \put(79,40){$ab = -ba$}
              \put(153,13){$-ab$}
              \put(184,106){$b$}
              \put(136,106){$a$}
              \put(188,77){\line(-5,3){46}}
              \put(142,80){\line(2,1){52}}
              \put(167,63){\line(3,2){21}}
              \put(165,65){\line(-3,2){23}}
              \put(167,63){\circle*{12}}
              \put(168,63){\line(0,-1){42}}
              \put(18,101){\line(1,-1){40}}
              \put(97,101){\line(-1,-1){39}}
              \put(58,62){\line(0,-1){42}}
              \put(57,62){\circle*{12}}
              \put(10,103){$a$}
              \put(90,102){$b$}
              \put(44,13){$ab$}
\end{picture}}

\begin{center} {\bf Figure 3 -- Diagrammatic Multiplication}
\end{center} \bigbreak

In Figure 4 we illustrate the Jacobi identity in the form $$(ac)b = (ab)c - a(bc).$$
\bigbreak

{\tt    \setlength{\unitlength}{0.92pt}
\begin{picture}(376,107)
\thinlines    \put(300,13){$D''$}
              \put(170,15){$D'$}
              \put(54,18){$D$}
              \put(69,89){$c$}
              \put(186,81){$c$}
              \put(156,81){$b$}
              \put(284,79){$b$}
              \put(134,33){$a$}
              \put(260,30){$a$}
              \put(327,82){$c$}
              \put(35,88){$b$}
              \put(10,35){$a$}
              \put(234,61){$-$}
              \put(99,62){$=$}
              \put(314,53){\circle*{12}}
              \put(314,54){\line(0,-1){27}}
              \put(314,53){\line(4,5){20}}
              \put(295,78){\line(4,-5){19}}
              \put(195,78){\line(0,-1){49}}
              \put(167,77){\line(0,-1){47}}
              \put(78,88){\line(-1,-2){28}}
              \put(45,86){\line(2,-3){35}}
\thicklines   \put(140,30){\vector(1,0){102}}
              \put(264,28){\vector(1,0){102}}
              \put(15,32){\vector(1,0){102}}
\end{picture}}

\begin{center} {\bf Figure 4 -- Diagrammatic Jacobi Identity}
\end{center} \bigbreak

To illustrate how this pattern can occur in a different context, consider diagrams $D$ of intersecting chords on a circle as shown
in Figure 5. By a circle we mean a curve in the plane without self-intersections that is a topological circle. By a chord, we mean
an arc without self-intersections that is embedded in the interior of the circle, touching the circle in two distinct points. Let us
suppose that we wish to color the chords from a set of
$q$ colors such that {\em if two chords intersect in an odd number of points, then  they receive different colors.} Let ${\cal
C}(D,q)$ denote the number of distinct colorings of the chords of the diagram $D$, as a  function of $q.$ Call such a diagram of
intersecting chords an {\em intersection graph}. We extend such diagrams by allowing internal trivalent vertices as illustrated in the
abstract by diagram $D''$ in Figure 4 and by the diagram with the same label, $D''$, in Figure 5. Interpret the trivalent vertex as an
instruction that {\em all chord lines meeting at a trivalent vertex receive the same color.} The diagrammatic Jacobi identity 
of Figure 4 corresponds directly to the logical coloring identity that says that if we have three diagrams $D,D',D''$ with
two chords touching in an odd number of points in $D$, one point removed in $D'$, and the two chords fused by a trivalent vertex
in $D''$ so that they must receive the same color, then {\em the number of colorings of $D$ is the number of colorings of $D'$ minus
the number of colorings of $D''$.} This is just the coloring version of the logical identity 
$$ Different \,  = \,  Anything \,  - \, Same.$$ For graph coloring problems, this identity was first articulated by 
Hassler Whitney \cite{Whitney}. In formulas, we have
$${\cal C}(D,q) = {\cal C}(D',q) - {\cal C}(D'',q).$$

{\tt    \setlength{\unitlength}{0.92pt}
\begin{picture}(385,144)
\thinlines    \put(297,13){$D''$}
              \put(171,15){$D'$}
              \put(49,16){$D$}
              \put(314,69){\circle*{12}}
              \put(314,68){\line(0,-1){37}}
              \put(314,67){\line(1,3){21}}
              \put(297,132){\line(1,-4){16}}
              \put(319,81){\oval(112,102)}
              \put(263,86){\line(4,1){111}}
              \put(192,134){\line(0,-1){101}}
              \put(170,134){\line(0,-1){102}}
              \put(192,83){\oval(112,102)}
              \put(135,83){\line(4,1){111}}
              \put(82,134){\line(-1,-4){25}}
              \put(42,133){\line(1,-3){33}}
              \put(10,83){\line(4,1){111}}
              \put(66,83){\oval(112,102)}
\end{picture}}

\begin{center} {\bf Figure 5 -- Intersection Graphs}
\end{center} \bigbreak
 
The convention that we have adopted here -- that two chords are colored differently if and only if they intersect in an odd number
of points, makes a demand on the interpretation of the trivalent nodes. All arcs entering a given node must receive the same color.
After more nodes are added we will have connected components of the resulting graph that contain nodes (the outer circle is not
regarded as part of the graph). Call such a connected component a {\em web} in a given diagram. Each web is colored by a single 
color. We regard a chord without nodes as a (degenerate) web. We take the convention that if the total number of intersections  
between two distinct webs is odd, then they must receive different colors. Of course, a web may have self-intersections; we define the
sign of the coloring of a given web to be $-1$ if it has an odd number of self-intersections and $+1$ if it has an even number of 
self-intersections. The sign of the coloring of a diagram is the product of the signs of its component webs. Note the the sign of a
chord is positive. With these conventions, the formulas in Figures 4 and 5 match perfectly and can be understood as indicating parts
of larger diagrams that differ only as indicated. We see, as in Figure 6, that an extra  self-intersection added to a trivalent
vertex changes the sign of its web. This corresponds to the algebraic interpretation of such as vertex as $ab=-ba$. See Figure 3.
\bigbreak

{\tt    \setlength{\unitlength}{0.92pt}
\begin{picture}(375,417)
\thinlines    \put(162,57){$=$}
              \put(310,24){\line(1,0){1}}
              \put(278,62){\line(0,1){27}}
              \put(239,62){\line(0,1){27}}
\thicklines   \put(208,12){\vector(1,0){102}}
\thinlines    \put(239,62){\line(4,-5){19}}
              \put(258,37){\line(4,5){20}}
              \put(258,38){\line(0,-1){27}}
              \put(258,37){\circle*{12}}
              \put(123,39){\circle*{12}}
              \put(123,40){\line(0,-1){27}}
              \put(123,39){\line(4,5){20}}
              \put(104,64){\line(4,-5){19}}
\thicklines   \put(73,14){\vector(1,0){102}}
\thinlines    \put(104,65){\line(1,1){33}}
              \put(142,65){\line(-1,1){30}}
              \put(245,158){$+$}
              \put(332,164){\line(-1,1){30}}
              \put(294,164){\line(1,1){33}}
              \put(71,162){\line(-1,1){32}}
              \put(43,162){\line(5,6){28}}
\thicklines   \put(137,115){\vector(1,0){102}}
              \put(263,113){\vector(1,0){102}}
              \put(16,115){\vector(1,0){102}}
\thinlines    \put(43,162){\line(0,-1){47}}
              \put(71,163){\line(0,-1){49}}
              \put(294,163){\line(4,-5){19}}
              \put(313,138){\line(4,5){20}}
              \put(313,139){\line(0,-1){27}}
              \put(313,138){\circle*{12}}
              \put(107,153){$=$}
              \put(211,57){$-$}
              \put(170,115){\line(0,1){70}}
              \put(197,115){\line(0,1){70}}
              \put(71,218){\line(0,1){70}}
              \put(44,218){\line(0,1){70}}
              \put(230,247){$-$}
              \put(95,248){$=$}
              \put(310,239){\circle*{12}}
              \put(310,240){\line(0,-1){27}}
              \put(310,239){\line(4,5){20}}
              \put(291,264){\line(4,-5){19}}
              \put(191,264){\line(0,-1){49}}
              \put(163,263){\line(0,-1){47}}
\thicklines   \put(136,216){\vector(1,0){102}}
              \put(260,214){\vector(1,0){102}}
              \put(11,218){\vector(1,0){102}}
\thinlines    \put(163,263){\line(5,6){28}}
              \put(191,263){\line(-1,1){32}}
              \put(291,265){\line(1,1){33}}
              \put(329,265){\line(-1,1){30}}
              \put(328,365){\line(-1,1){30}}
              \put(290,365){\line(1,1){33}}
              \put(190,363){\line(-1,1){32}}
              \put(162,363){\line(5,6){28}}
              \put(73,374){\line(-1,1){30}}
              \put(40,372){\line(1,1){34}}
\thicklines   \put(10,318){\vector(1,0){102}}
              \put(259,314){\vector(1,0){102}}
              \put(135,316){\vector(1,0){102}}
\thinlines    \put(40,372){\line(2,-3){35}}
              \put(73,374){\line(-1,-2){28}}
              \put(162,363){\line(0,-1){47}}
              \put(190,364){\line(0,-1){49}}
              \put(290,364){\line(4,-5){19}}
              \put(309,339){\line(4,5){20}}
              \put(309,340){\line(0,-1){27}}
              \put(309,339){\circle*{12}}
              \put(94,348){$=$}
              \put(229,347){$-$}
\end{picture}}

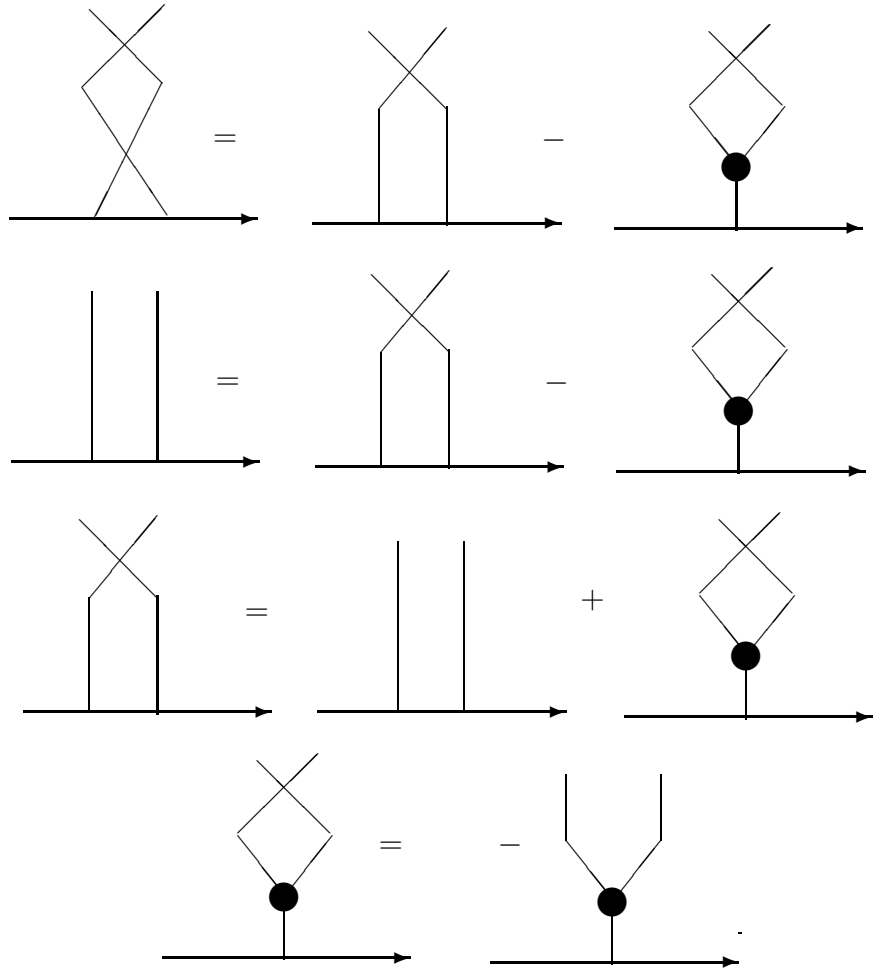
\begin{center} {\bf Figure 6 -- Verifying the Twist Identity for Color Diagrams}
\end{center} \bigbreak

In Figure 6 we illustrate how these sign conventions are consistent with the coloring formula/Jacobi identity.
In this figure, we begin with the Jacobi identity with a twist (crossing) added to each diagram. The original diagram with one crossing
now  has two, and hence is equivalent to a diagram with none (no local requirement of difference). The original diagram with no
crossing now has one, and is interpreted as a requirement of difference. Rearranging, we find the Jacobi identity again, but with an
extra crossing and change of sign for the noded diagram. The conclusion is that adding a crossing to a node changes the sign of its
diagram.
\bigbreak

We see that the patterns of counting colorings of chord diagrams correspond formally to the axioms for a Lie algebra. This example
indicates how a combinatorial context can lead to the very formalism on which this paper is based, but though different structures
than one could have initially visualized. Diagrammatic Lie algebras similar to this example feature prominently in the theory of 
Vassiliev invariants \cite{Bar-Natan, Lieberum} of knots and links, and may form the basis for new models for the structures that we
have discussed in this paper.
\bigbreak

We have concentrated in this section on a coloring example not only because the occurrence of the Jacobi identity in this context may
appear startling, but also because there is a more direct relationship with coloring in regard to the fundamental Lie algebra of
$SU(2)$ (or equivalently $SO(3)$) that underlies the structures we have discussed in this paper. The Lie algebra of $SO(3)$ has
structure constant the alternating epsilon symbol $\epsilon_{ijk}$ that we have used again and again in Section 7 for the
generalization of the Feynman-Dyson derivation. This epsilon can be expressed diagrammatically as a trivalent vertex.
The basic epsilon identity $$\epsilon_{abi}\epsilon_{cdi} =  -\delta_{ad}\delta_{bc} + \delta_{ac}\delta_{bd}$$ can be written 
diagrammatically, and it leads at once to a diagrammatic Jacobi identity. See Figure 7 for the diagrammatic form of the epsilon
identity. The epsilon itself is closely related to coloring (See \cite{Pen:Spin,Penrose, Kauff:KP}), but that is another story and we
shall stop here.
\bigbreak

{\tt    \setlength{\unitlength}{0.92pt}
\begin{picture}(345,362)
\thinlines    \put(98,287){$\epsilon_{abi}$}
\thicklines   \put(41,239){$i$}
              \put(80,352){$b$}
              \put(8,187){$a$}
              \put(217,123){$-$}
              \put(100,118){$=$}
              \put(126,55){$d$}
              \put(244,55){$d$}
              \put(323,54){$c$}
              \put(202,56){$c$}
              \put(321,188){$b$}
              \put(200,188){$b$}
              \put(241,188){$a$}
              \put(125,188){$a$}
              \put(56,121){$i$}
              \put(1,353){$a$}
              \put(3,52){$d$}
              \put(80,53){$c$}
              \put(82,189){$b$}
              \put(74,4){$\epsilon_{abi}\epsilon_{cdi} = \delta_{ac}\delta_{bd}
               - \delta_{ad}\delta_{bc}$}
              \put(211,185){\line(-2,-3){80}}
              \put(132,185){\line(2,-3){78}}
              \put(311,83){\line(1,-1){19}}
              \put(311,163){\line(0,-1){81}}
              \put(329,183){\line(-1,-1){18}}
              \put(270,84){\line(-1,-1){19}}
              \put(270,164){\line(0,-1){79}}
              \put(250,183){\line(1,-1){19}}
              \put(52,105){\line(-1,-1){41}}
              \put(52,104){\line(1,-1){39}}
              \put(52,144){\line(0,-1){40}}
              \put(53,104){\circle*{18}}
              \put(53,144){\circle*{18}}
              \put(12,184){\line(1,-1){38}}
              \put(91,184){\line(-1,-1){35}}
              \put(51,312){\line(0,-1){62}}
              \put(91,349){\line(-1,-1){35}}
              \put(12,349){\line(1,-1){38}}
              \put(53,309){\circle*{18}}
\end{picture}}

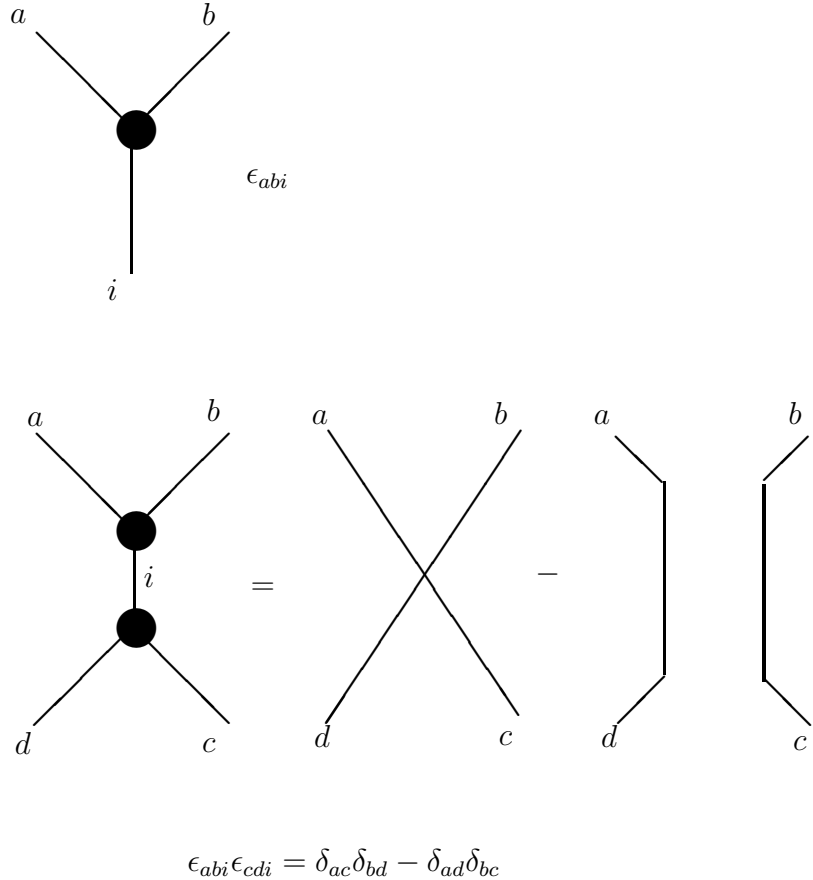
\begin{center} {\bf Figure 7 -- The Epsilon Identity in Diagrammatic Form.}
\end{center} \bigbreak

\section{Epilogue}
We have sought in this paper, to begin in an algebraic framework that naturally contains the formalism of the
calculus, but not its notions of limits or constructions of spaces with specific locations, points and trajectories. It is
remarkable that so many patterns of  physical law fit so well such an abstract framework. We believe that this is indicative of the
secondary nature of point sets, topologies and classical differential geometries in physics (Compare \cite{Crane}). In this paper we
have dispensed with spacetime and replaced it by algebraic structure. But behind that structure, the space stands ready to be
constructed, by discrete derivatives and patterns of steps, or by starting with a discrete pattern in the form of a diagram, a
network, a lattice, a knot, or a simplicial complex, and elaborating that structure until  the specificity of spatio-temporal locations
appear.
\bigbreak 

There are many ideas for producing location. Poisson brackets allow us to connect
classical notions of location with the non-commutative algebra used herein.  Below the level of the Poisson brakets is a treatment of 
processes and operators as though they were variables in the same context as the variables in the classical calculus.
In different degrees we have let go of the notion of classical variables and yet retained their form, as we made a descent into the 
discrete. 
\bigbreak

In order for locations to appear from process, one may need an appropriate degree of recursiveness. Lie
algebras begin the process with their fully self-operant structure of derivations. It is just  such bootstrapping that fits into
the basis of our concerns and produces the ways to make spaces emerge, through process, from abstract algebra.
\bigbreak

\end{document}